\documentclass[twocolumn]{aastex61}
\pdfoutput=1 
\usepackage{amsmath,amstext}
\usepackage[T1]{fontenc}
\usepackage{apjfonts} 
\usepackage[figure,figure*]{hypcap}

\shorttitle{The Dust-Selected Molecular Clouds in the SMC}
\shortauthors{Takekoshi et al.}

\begin{document}

\title{The Dust-Selected Molecular Clouds in the Northeast Region of the Small Magellanic Cloud\footnote{\textit{Herschel} is an ESA space observatory with science instruments provided by European-led Principal Investigator consortia and with important participation from NASA.}}

\author[0000-0002-4124-797X]{Tatsuya Takekoshi}
\affiliation{Institute of Astronomy, The University of Tokyo, 2-21-1 Osawa, Mitaka, Tokyo 181-0015, Japan}
\affiliation{Graduate School of Informatics and Engineering, The University of Electro-Communications, Chofu, Tokyo 182-8585, Japan}

\author[0000-0001-9778-6692]{Tetsuhiro Minamidani}
\affiliation{Nobeyama Radio Observatory, National Astronomical Observatory of Japan (NAOJ), National Institutes of Natural Sciences (NINS), 462-2, Nobeyama, Minamimaki, Minamisaku, Nagano 384-1305, Japan}
\affiliation{Department of Astronomical Science, School of Physical Science, SOKENDAI (The Graduate University for Advanced Studies), 2-21-1, Osawa, Mitaka, Tokyo 181-8588, Japan}

\author{Shinya Komugi}
\affiliation{Kogakuin University, 2665-1 Nakano, Hachioji, Tokyo 192-0015, Japan}

\author[0000-0002-4052-2394]{Kotaro Kohno}
\affiliation{Institute of Astronomy, The University of Tokyo, 2-21-1 Osawa, Mitaka, Tokyo 181-0015, Japan}
\author[0000-0001-9016-2641]{Tomoka Tosaki}
\affiliation{Joetsu University of Education, Joetsu, Niigata 943-8512, Japan}
\author{Kazuo Sorai}
\affiliation{Department of Physics, Faculty of Science, Hokkaido University, Sapporo 060-0810, Japan}
\author{Erik Muller}
\affiliation{Chile Observatory, National Astronomical Observatory of Japan (NAOJ), National Institutes of Natural Sciences (NINS), 2-21-1, Osawa, Mitaka, Tokyo 181-8588, Japan}

\author{Norikazu Mizuno}
\affiliation{Chile Observatory, National Astronomical Observatory of Japan (NAOJ), National Institutes of Natural Sciences (NINS), 2-21-1, Osawa, Mitaka, Tokyo 181-8588, Japan}
\affiliation{Department of Astronomical Science, School of Physical Science, SOKENDAI (The Graduate University for Advanced Studies), 2-21-1, Osawa, Mitaka, Tokyo 181-8588, Japan}
\affiliation{Joint ALMA Observatory, Alonso de C\'ordova 3107, Vitacura, Santiago 763-0355, Chile.}

\author{Akiko Kawamura}
\affiliation{Chile Observatory, National Astronomical Observatory of Japan (NAOJ), National Institutes of Natural Sciences (NINS), 2-21-1, Osawa, Mitaka, Tokyo 181-8588, Japan}

\author{Toshikazu Onishi}
\affiliation{Department of Physical Science, Osaka Prefecture University, Gakuen 1-1, Sakai, Osaka 599-8531, Japan}

\author{Yasuo Fukui}
\affiliation{Department of Astrophysics, Nagoya University, Chikusa-ku, Nagoya 464-8602, Japan}

\author{Caroline Bot}
\affiliation{Universit\'{e} de Strasbourg, CNRS, Observatoire astronomique de Strasbourg, UMR 7550, F-67000 Strasbourg, France}

\author{Monica Rubio}
\affiliation{Departamento de Astronom\'{i}a, Universidad de Chile, Casilla 36-D, 8320000 Santiago, Chile}

\author{Hajime Ezawa}
\affiliation{Chile Observatory, National Astronomical Observatory of Japan (NAOJ), National Institutes of Natural Sciences (NINS), 2-21-1, Osawa, Mitaka, Tokyo 181-8588, Japan}
\affiliation{Department of Astronomical Science, School of Physical Science, SOKENDAI (The Graduate University for Advanced Studies), 2-21-1, Osawa, Mitaka, Tokyo 181-8588, Japan}

\author{Tai Oshima}
\affiliation{Advanced Technology Center, National Astronomical Observatory (NAOJ), National Institutes of Natural Sciences (NINS), 2-21-1, Osawa, Mitaka, Tokyo 181-8588, Japan}
\affiliation{Department of Astronomical Science, School of Physical Science, SOKENDAI (The Graduate University for Advanced Studies), 2-21-1, Osawa, Mitaka, Tokyo 181-8588, Japan}

\author{Jason E. Austermann}
\affiliation{National Institute of Standards and Technology, Boulder, CO 80305, USA}

\author{Hiroshi Matsuo}
\affiliation{Advanced Technology Center, National Astronomical Observatory (NAOJ), National Institutes of Natural Sciences (NINS), 2-21-1, Osawa, Mitaka, Tokyo 181-8588, Japan}
\affiliation{Department of Astronomical Science, School of Physical Science, SOKENDAI (The Graduate University for Advanced Studies), 2-21-1, Osawa, Mitaka, Tokyo 181-8588, Japan}

\author{Itziar Aretxaga}
\affiliation{Instituto Nacional de Astrof\'{i}sica, \'{O}ptica y Electr\'{o}nica (INAOE), 72000 Puebla, Mexico}

\author{David H. Hughes}
\affiliation{Instituto Nacional de Astrof\'{i}sica, \'{O}ptica y Electr\'{o}nica (INAOE), 72000 Puebla, Mexico}

\author{Ryohei Kawabe}
\affiliation{Division of Radio Astronomy, National Astronomical Observatory of Japan (NAOJ), National Institutes of Natural Sciences (NINS), 2-21-1, Osawa, Mitaka, Tokyo 181-8588, Japan}
\affiliation{Department of Astronomical Science, School of Physical Science, SOKENDAI (The Graduate University for Advanced Studies), 2-21-1, Osawa, Mitaka, Tokyo 181-8588, Japan}

\author{Grant W. Wilson}
\affiliation{Department of Astronomy, University of Massachusetts, Amherst, MA 01003, USA}

\author{Min S. Yun}
\affiliation{Department of Astronomy, University of Massachusetts, Amherst, MA 01003, USA}

\begin{abstract}
We present a high-sensitivity ($1\sigma<1.6~\mathrm{mJy~beam^{-1}}$) continuum observation in a 343~arcmin$^2$ area of the northeast region in the Small Magellanic Cloud at a wavelength of 1.1~mm, conducted using the AzTEC instrument on the ASTE telescope.
In the observed region, we identified 20 objects by contouring $10\sigma$ emission.
Through spectral energy distribution (SED) analysis using 1.1~mm, \textit{Herschel}, and \textit{Spitzer} data, we estimated the gas masses of $5\times 10^3$--$7\times 10^4~\mathrm{M_\odot}$, assuming a gas-to-dust ratio of 1000.
Dust temperature and the index of emissivity were also estimated as 18--33~K and 0.9--1.9, respectively, which are consistent with previous low resolution studies.
The relation between dust temperature and the index of emissivity shows a weak negative linear correlation.
We also investigated five CO-detected dust-selected clouds in detail. The total gas masses were comparable to those estimated from the Mopra CO data, indicating that the assumed gas-to-dust ratio of 1000 and the $X_\mathrm{CO}$ factor of $1\times10^{21}~\mathrm{cm^{-2}~(K~km~s^{-1})^{-1}}$, with uncertainties of a factor of 2, are reliable for the estimation of the gas masses of molecular or dust-selected clouds.
Dust column density showed good spatial correlation with CO emission, except for an object that associates with bright young stellar objects.
The $8~\micron$ filamentary and clumpy structures also showed similar spatial distribution with the CO emission and dust column density, supporting the fact that polycyclic aromatic hydrocarbon emissions arise from the surfaces of dense gas and dust clouds.
\end{abstract}


\section{Introduction}
The Small Magellanic Cloud (SMC) is a dwarf galaxy characterized by a metal-poor environment \citep{1999ApJ...518..246K, 2000A&A...364..455L,2007ApJ...658.1027L,2011A&A...536A..17P,2014ApJ...797...85G} and active star formation \citep[e.g.,][]{1980A&A....90...73V,1993A&A...280..365L,2007ApJ...655..212B}.
Because of the proximity \citep[$\sim$60~kpc, e.g.,][]{2005MNRAS.357..304H} compared to other nearby galaxies, the SMC provides an invaluable opportunity to investigate the physics of the interstellar medium (ISM) and star formation along with the Large Magellanic Cloud (LMC) \citep[e.g.,][]{2010ARA&A..48..547F}.

Previously, studies to unveil the star formation activity in the SMC were primarily motivated by the detection of giant molecular clouds (GMCs), which are the principal formation sites of stars \citep[e.g.,][]{1993A&A...271....1R, 1993A&A...271....9R, 1996A&AS..118..263R, 2000A&A...359.1139R, 1994A&A...292..371L, 1993A&A...276...25I, 2003A&A...406..817I, 2003MNRAS.338..609M, 2015MNRAS.448.1847H}.
The first GMC survey toward the full SMC was conducted by the Columbia survey, where the detection of five GMCs detection was reported, with the masses of $\sim 10^6$--$10^7~M_\sun$, at $\sim$ 160 pc resolution \citep{1991ApJ...368..173R}.
A subsequent GMC survey at $\sim$ 50 pc resolution was conducted with the NANTEN 4-m telescope, and twenty-one GMCs, whose masses were $\sim 10^4$--$10^6~M_\sun$, were detected \citep{2001PASJ...53L..45M}.
High-resolution follow-up observations toward the NANTEN GMCs were also conducted by the Mopra telescope \citep{2010ApJ...712.1248M, 2013IAUS..292..110M}, resolving the NANTEN GMCs into compact ($10^3$--$10^4~M_\sun$) molecular clumps.
These studies pointed out that CO emission in the SMC is very weak, and the conversion factor, $X_\mathrm{CO}\simeq 10^{21}~\mathrm{cm^{-2} (K~km~s^{-1})^{-1}}$, is about 10 times larger than the typical value in the Milky Way galaxy.
Recent numerical studies have suggested that the formation of molecules ($\mathrm{H_2}$ and CO) in the low-metallicity ISM can occur later than the beginning of star formation \citep{2012MNRAS.421....9G,2012MNRAS.426..377G}.
These results imply that observations of CO lines are not always suitable for the detection of dense gas clouds that are connected to star formation in low-metallicity environments.

As a complementary approach to CO line observations, dust continuum emission can also be a good tracer of dense gas clouds via thermal emission from cold dust.
Toward the SMC southwest (SW) region, observations by SEST/SIMBA at 1.2~mm \citep{2004A&A...425L...1R,2007A&A...471..103B} and APEX/LABOCA at 870~$\micron$ succeeded at detecting the CO clouds \citep{2010A&A...524A..52B}.
N66 has also been investigated by LABOCA and \textit{Herschel} in detail by \citet{2015MNRAS.448.1847H}.
Recently, \citet{2017ApJ...835...55T} attempted a new cloud identification method using 1.1~mm continuum survey data toward the full SMC obtained by AzTEC on ASTE.
They identified 44 dust-selected cloud samples in the full SMC with a gas mass range of $4\times 10^3$--$3\times 10^5 M_\sun$.
This survey also discovered dust-selected clouds not associated with CO lines or star formation activity; these objects are expected to be rare samples of the youngest evolution phase of GMCs in the SMC.

\citet{2017ApJ...835...55T}, however, did not report the detection of the two CO clouds discovered by NANTEN in the northeast (NE) region.
These objects are characterized by relatively weak star formation activity compared to the other GMCs in the SMC.
This suggests that these clouds have a low dust temperature or low surface brightness, which are not sufficient to detect the peak flux, because of the inadequate survey sensitivity.
Thus, it is important to observe these objects by the 1.1~mm continuum with high-sensitivity observations.

In this paper, we present the results of a 1.1~mm deep observation toward the SMC NE region in order to reveal the hidden aspects of GMCs in the low-metallicity ISM using dust, CO, and star formation tracers.
In Section~\ref{sec2}, the observation and data reduction of the AzTEC and \textit{Herschel} data are described.
In Section~\ref{sec3}, we present the multi-wavelength images and catalog of the 1.1~mm objects.
Section~\ref{sec4} describes the method of spectral energy distribution (SED) analysis by both object- and image-based approaches.
The statistical properties of the low-mass dust-selected clouds in the NE region are discussed in Section~\ref{sec5}.
The relationship between the distribution of dust, CO, and star formation activity are discussed with the result of the image-based SED fitting toward the NANTEN GMCs in Section~\ref{sec6}.
Finally, we summarize the results of this study in Section~\ref{sec7}.

\section{Observation and datasets} 
\label{sec2}
\subsection{AzTEC/ASTE observation}
\begin{figure}
\epsscale{1.33}
\plotone{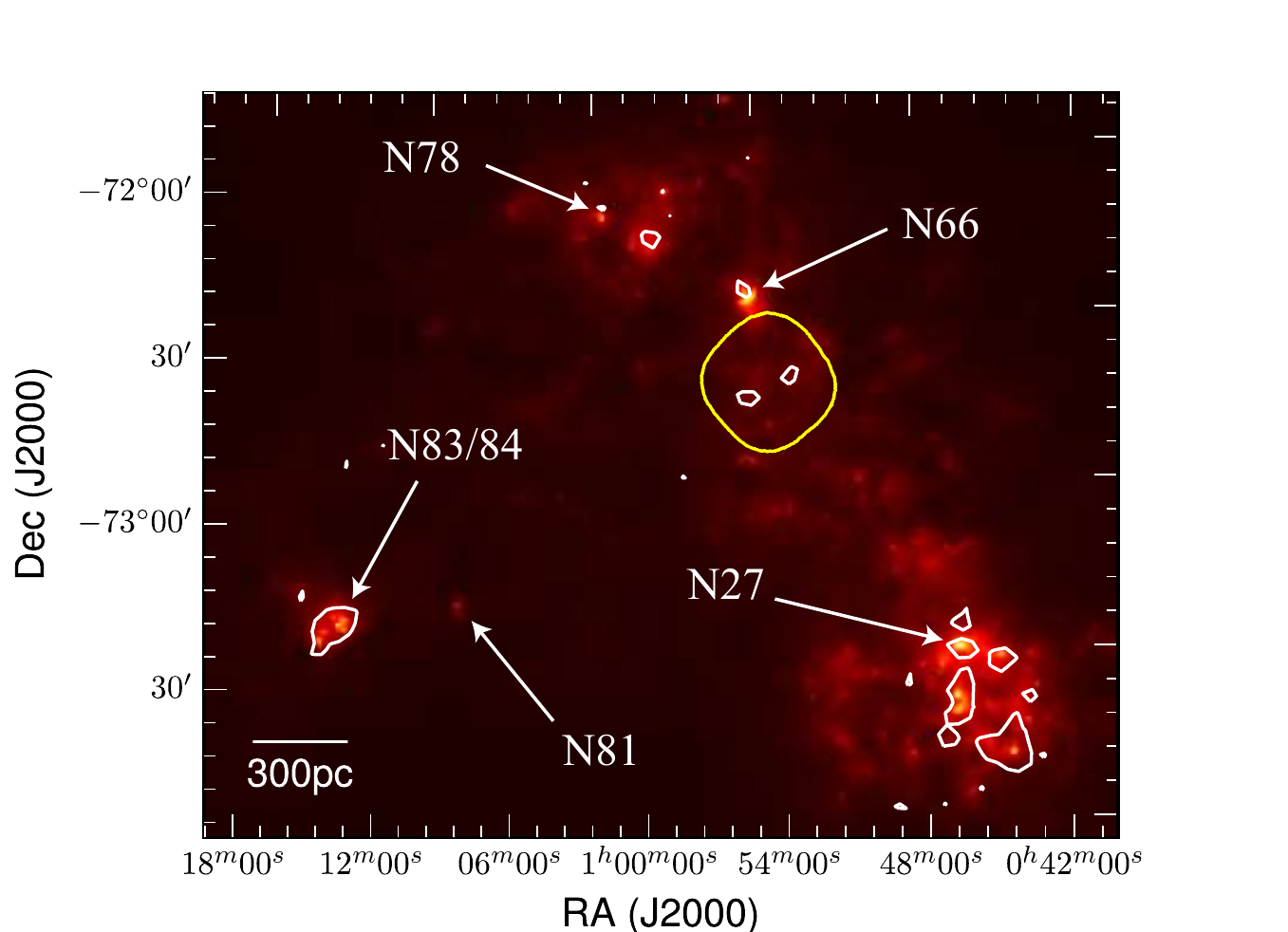}
\caption{\textit{Spitzer} $160~\micron$ image of the SMC. The observation regions of AzTEC/ASTE are represented by the yellow outline. The white contours represent the NANTEN CO intensity (0.5~$\mathrm{K~km~s^{-1}}$). \label{fig_obsregion}}
\end{figure}

Observations of the 1.1~mm continuum toward the SMC NE region were conducted by the AzTEC instrument \citep{2008MNRAS.386..807W} mounted on the ASTE 10~m telescope \citep{2004SPIE.5489..763E, ezawa2008new} on August 28--31, October 4--6, 2007, and August 28--30, 2008.
The observation region is shown in Figure~\ref{fig_obsregion}.
The zenith opacity at 220~GHz was in the range of 0.01--0.16 and the median was 0.06.
The total observing time was about 20 hours.
The observations were performed by $16\arcmin \times 16\arcmin$ and $10\arcmin \times 10\arcmin$ lissajous scans with a peak velocity of $330\arcsec~\mathrm{s^{-1}}$. 
Pointing observations toward quasars J0047-579 or 2355-534 were observed every 1.5~hours, and the accuracy was better than $3\arcsec$ rms \citep{2008MNRAS.390.1061W}.
The uncertainty of the flux calibration by Uranus was 8\% \citep{2008MNRAS.390.1061W, 2010AJ....139.1190L}.

Data reduction was performed by the FRUIT method, which is an iterative principal component analysis cleaning method \citep{2010AJ....139.1190L,2012MNRAS.423..529D, 2008MNRAS.385.2225S} to recover extended emission.
The FRUIT method effectively removes atmospheric emission, which is a dominant noise source of ground-based continuum observation, extended over the field of view of the instrument, and correlated among bolometer pixels. At the same time, widely extended astronomical signals, which are also correlated among bolometer pixels, are also removed.
Simulations of reproducibility of extended sources using Gaussian model sources show that $>60\%$ of the total flux density and $\sim 100\%$ of peak flux density are recovered for compact objects smaller than $<3'$ full-width at half-maximum (FWHM) in our cases, which corresponds to the typical size of the detected objects in the observation region, as shown in Section \ref{sec3} \citep{2011PASJ...63.1139K,2011PASJ...63..105S,2017ApJ...835...55T}.

As a result of the FRUIT method, the achieved minimum and median noise levels were $1.16$ and $1.31~\mathrm{mJy~beam^{-1}}$, respectively.
The total area better than $1.64~\mathrm{mJy~beam^{-1}}$, which is $\sqrt{2} \times$ the minimum noise level and used for further data analysis in this paper, was 343.4 square arcminutes (roughly $20'\times 20'$).
The FWHM of the point response function (PRF) after the FRUIT procedure was $40\arcsec$, which corresponds to 12~pc.
FRUIT also added an uncertainty of 10\% to the photometry of the detected objects \citep{2017ApJ...835...55T}.
The total photometric accuracy was estimated to be 13\% by the root sum square of the flux calibration and FRUIT photometric accuracy. 

\subsection{\textit{Herschel} data}
We used the \textit{Herschel}/PACS (100 and 160~$\micron$) and SPIRE (250, 350, and 500~$\micron$) data \citep{2013AJ....146...62M, 2014ApJ...797...85G} to estimate the amount and property of the cold dust component.
Some part of the extended component of the 1.1~mm data analyzed by the FRUIT procedure, which is removed as correlated noise similar to atmospheric emission, is lost.
In order to compare with the 1.1~mm data directly, the FRUIT procedure was applied to the \textit{Herschel} images in the same manner as \citet{2017ApJ...835...55T}.
After the FRUIT procedure, the FWHM of the PRF was $40\arcsec$.
The image noise levels after the FRUIT analysis were 383, 213, 25.6, 12.8, and 8.5~$\mathrm{mJy~beam^{-1}}$ for the \textit{Herschel} 100, 160, 250, 350, and 500~$\micron$ data, respectively. 
We also considered the propagation of 1.1~mm image noise that was caused by the FRUIT analysis.
The photometric accuracy of the \textit{Herschel} data after the FRUIT analysis was 14\% and 13\% for the PACS and SPIRE data, respectively, which was estimated by the root sum square of the calibration accuracy (8\% and 10\% for SPIRE and PACS data, respectively) and additional error by the FRUIT analysis \cite[10\%,][]{2017ApJ...835...55T}.

\subsection{Molecular gas and star formation tracers}
We used the CO $(J=1-0)$ data obtained by the Mopra telescope \citep[MAGMA-SMC,][]{2010ApJ...712.1248M, 2013IAUS..292..110M}.
The observation was made toward the GMCs detected by the NANTEN survey \citep{2001PASJ...53L..45M}. The spatial resolution was $33\arcsec$ FWHM with a sensitivity of about 150~mK and velocity resolution binned to 0.35~km~s$ ^{-1}$.
In the observation region of the AzTEC 1.1~mm continuum, \citet{2010ApJ...712.1248M} reported the detection of 4~CO clumps in this region, which have gas masses of $10^3$--$10^4~\mathrm{M_\sun}$ with the assumption of $X_\mathrm{CO} = 1\times 10^{21}~\mathrm{cm^{-1}~(K~km~s^{-1})^{-1}}$.

In order to investigate the star formation activity, we used the \textit{Sptizer}/IRAC $8~\micron$ and MIPS $24$ and $70~\micron$ data \citep{2011AJ....142..102G}. We did not apply FRUIT analysis, as the spatial distributions differed greatly from the AzTEC and \textit{Herschel} data because of the tracing of different dust components (very small grains or polycyclic aromatic hydrocarbons (PAHs)). The noise levels and photometric accuracy were 0.02, 0.06, and 0.5~MJy~sr$^{-1}$, and 5\%, 4\%, and 5\% for 8, 24, and $70~\micron$, respectively.
We also used the \textit{Spitzer} young stellar object (YSO) catalogs provided by \citet{2007ApJ...655..212B} and \citet{2013ApJ...778...15S} to determine whether star formation activity is associated.
In addition, we compared with the H$\alpha$ data obtained by the Magellanic Cloud Emission-Line Survey project \citep{2000ASPC..221...83S}, and radio continuum data at 8.64 and 4.8~GHz \citep[ATCA and Parkes,][]{2010AJ....140.1511D} as another tracer of star-forming regions.

\section{Results}
\label{sec3}
\subsection{Maps}
\label{sec31}
Figure~\ref{fig_maps} shows the continuum emission at AzTEC 1.1~mm, \textit{Herschel} $100$--$500~\micron$, and \textit{Spitzer} $8$, $24$, and $70~\micron$ in the observation region.
We did not subtract free-free emissions from the maps, because the observed region shows no strong radio continuum emission.
The 1.1~mm emission over $5\sigma$ shows good spatial correlation with the \textit{Herschel} images. 
This indicates that the AzTEC and \textit{Herschel} bands trace a cold dust component.
In contrast, the images at shorter wavelengths exhibit a compact spatial distribution because these bands efficiently trace thermal dust emission from hotter and more compact star formation activity \citep[e.g.,][]{2007ApJ...655..212B,2007ApJ...658.1027L}.
All CO clouds discovered by NANTEN and Mopra in this region were detected over 10$\sigma$ in the 1.1~mm image, which we will discuss in Section~\ref{sec6}.

\begin{figure*}
\vspace{-1cm}
\plotone{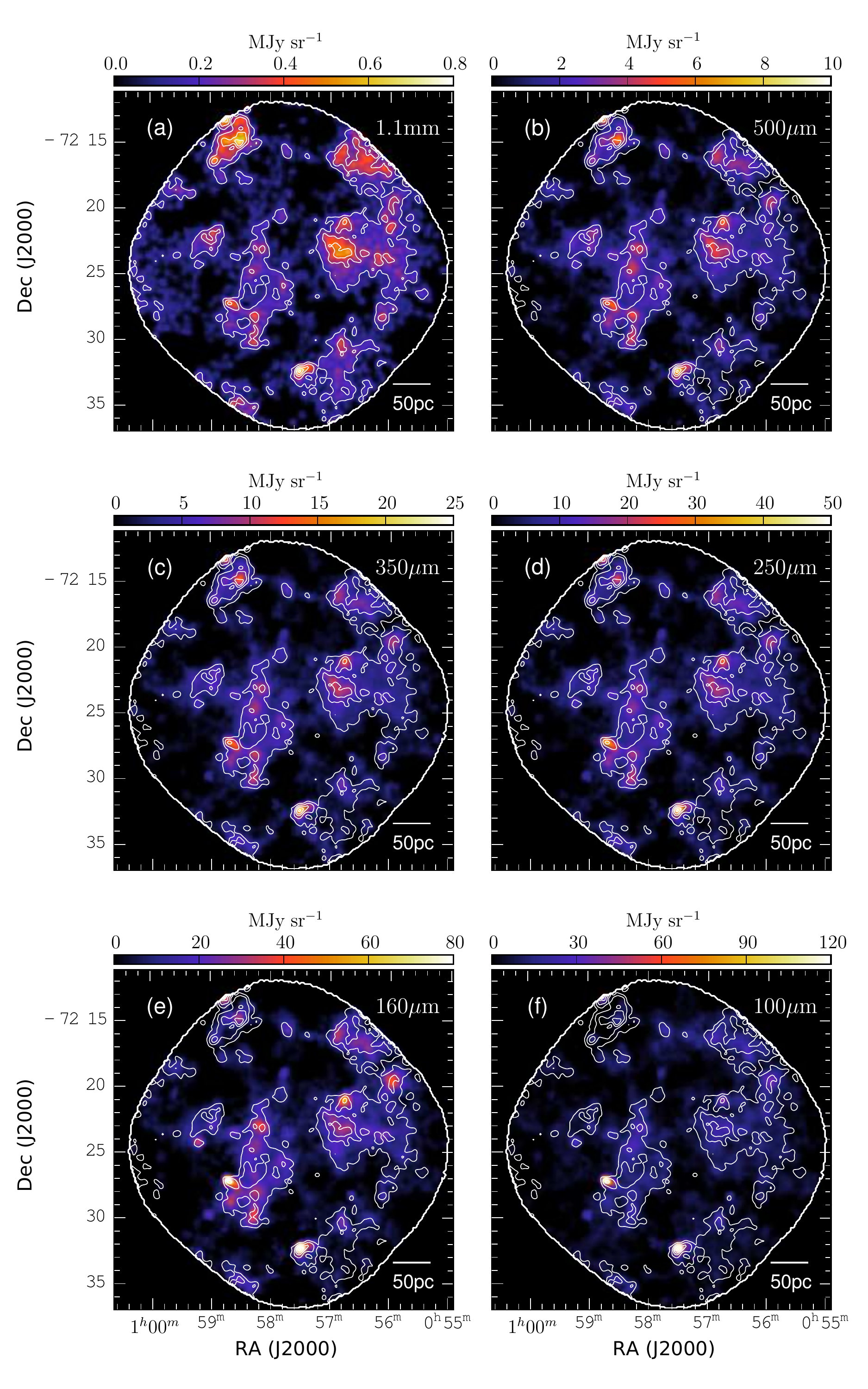}
\caption{Multi-band continuum images of the SMC NE region. The white contours represent the 1.1~mm image, starting from 6.5~mJy~$\mathrm{beam^{-1}}$ ($\sim 5\sigma$) with a step of 6.5~mJy~$\mathrm{beam^{-1}}$. (a) AzTEC/ASTE 1.1~mm. (b) \textit{Herschel}/SPIRE $500~\micron$. (c) \textit{Herschel}/SPIRE $350~\micron$. (d) \textit{Herschel}/SPIRE $250~\micron$. (e) \textit{Herschel}/PACS $160~\micron$. (f) \textit{Herschel}/PACS $100~\micron$. (g) \textit{Spitzer}/MIPS $70~\micron$. (h) \textit{Spitzer}/MIPS $24~\micron$. (i) \textit{Spitzer}/IRAC $8~\micron$. The edge of the 1.1~mm image is also indicated by the white contour.\label{fig_maps}}
\end{figure*}

\addtocounter{figure}{-1}
\begin{figure*}
\plotone{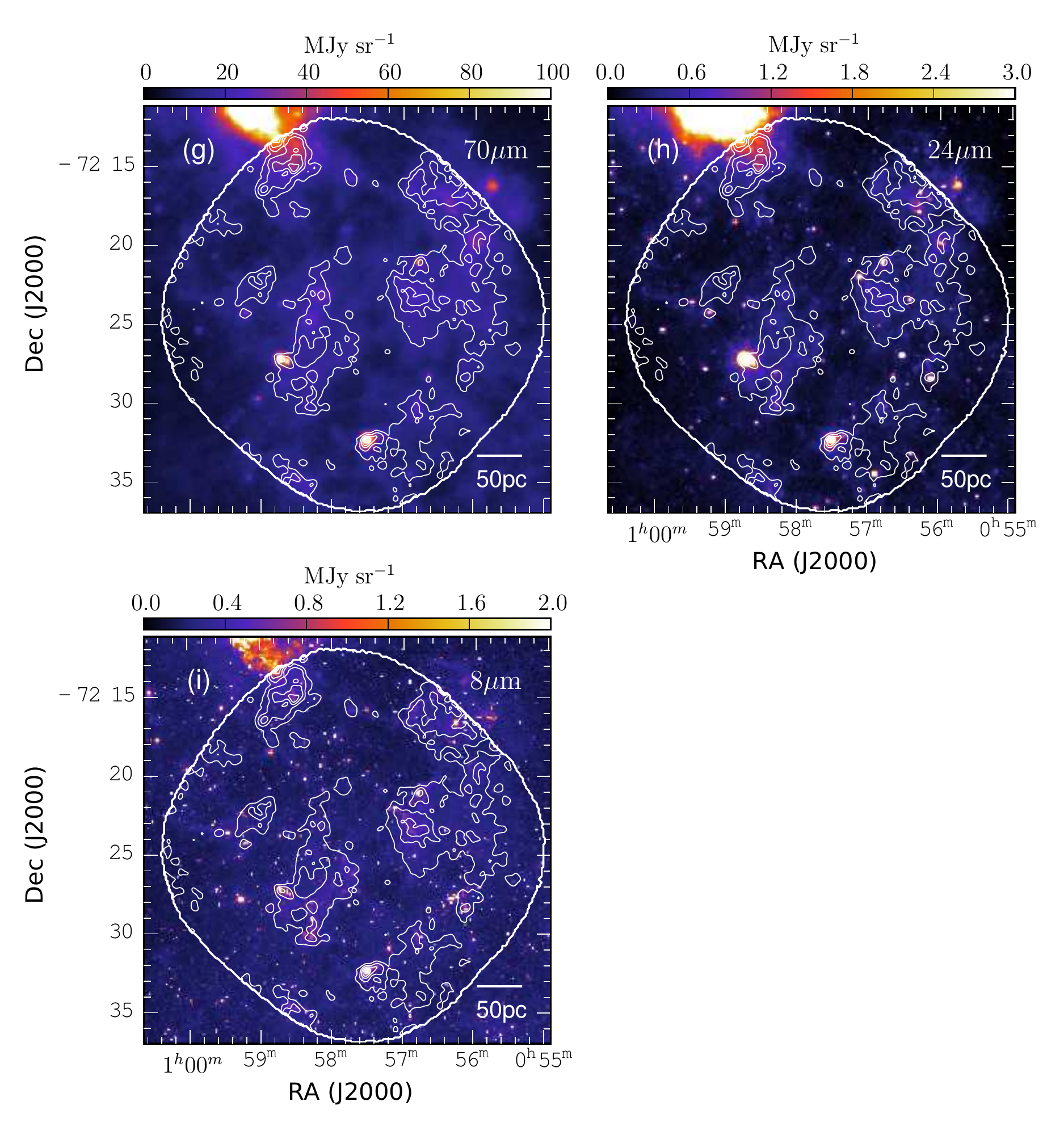}
\caption{(Continued.)}
\end{figure*}

\subsection{The 1.1~mm Object Catalog}
\label{sec32}
The 1.1~mm objects (hereafter called NEdeep objects) were identified by contouring over $10\sigma$ emission on the 1.1~mm image.
As a result, 20 objects, listed in Table~\ref{table:catalog_fruit}, were identified in the observing area.
Out of the detected objects, three objects were detected on the edge of the 1.1~mm image.
We did not use these objects for statistical analysis in Section~\ref{sec5}.
In the same manner as \citet{2017ApJ...835...55T}, the 1.1~mm objects were classified by whether star formation tracers such as H{\small II} regions, YSO candidates, or bright $24~\micron$ sources ($>10~\mathrm{MJy~sr^{-1}}$) exist in the objects.
As a result, we identified eight objects exhibiting star formation activity, and some of them are compact and faint at 1.1~mm such as NEdeep-15 and 18.
Therefore, the 1.1~mm objects are good candidates of dense gas clouds that are likely to form massive stars.

\begin{figure}
\epsscale{1.2}
\plotone{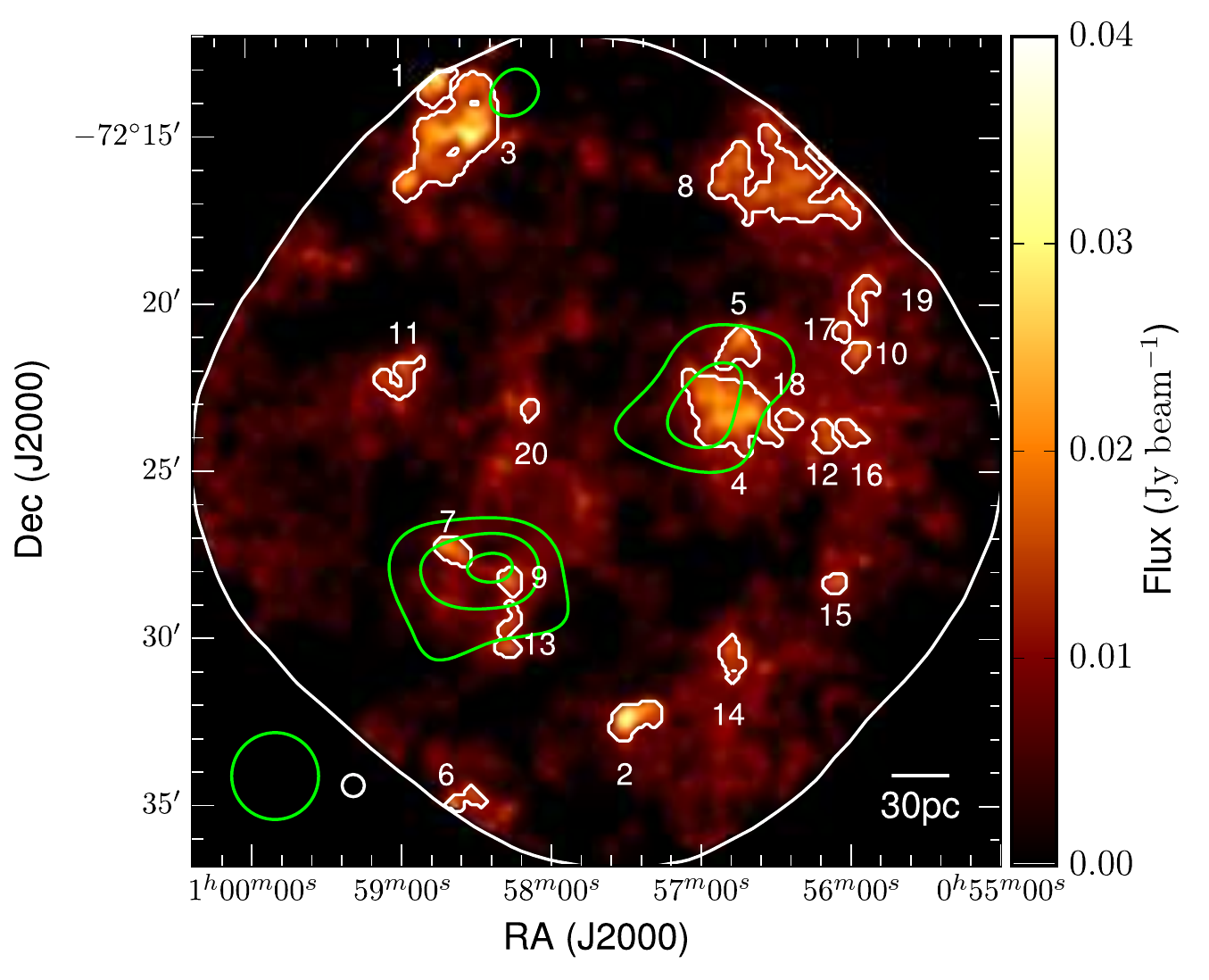}
\caption{The 1.1~mm image of the SMC NE region. The identified dust-selected clouds are shown in the white contours. The green contours represent the NANTEN CO intensity with a step of 0.3~$\mathrm{K~km~s^{-1}}$ starting from 0.3~$\mathrm{K~km~s^{-1}}$. The numbers represent the object IDs. The circles at the bottom left corner represent the effective resolutions of AzTEC/ASTE 1.1~mm (white) and NANTEN CO (green). \label{fig_aztec}}
\end{figure}

The spatial distribution of the identified NEdeep objects is presented in Figure~\ref{fig_aztec}.
The NANTEN objects at the eastern and western sides consist of two and three compact NEdeep objects, respectively. 
NEdeep-1 and 3, located at the northeast edge, are continuously connected from the N66 star-forming region.
The compact objects at the western side seem to be packed into some small regions and could be associated with each other. 
In contrast, NEdeep-2, 6, 11, 14, 15, and 20 seem to exist as independent compact objects.

\begin{deluxetable*}{rcccccccc}
\tabletypesize{\scriptsize}
\tablecaption{AzTEC/ASTE 1.1 mm extended source catalog of the SMC NE region. \label{table:catalog_fruit}}
\tablehead{Object ID&$\alpha$&$\delta$&1.1mm Peak flux&S/N&1.1 mm total flux&$R$&Star formation?&Map edge?\\
&$(J2000)$&$(J2000)$&(mJy~beam$^{-1}$)&&(mJy)&(pc)}
\startdata
NEdeep-1&00$^h$58$^m$45.9$^s$&-72$^\circ$13$'$18$''$&48.7 $\pm$2.3 &21.0 &28.8 $\pm$4.1 &8.0 &Yes&Yes\\
2&00$^h$57$^m$30.7$^s$&-72$^\circ$32$'$30$''$&34.1 $\pm$1.3 &27.2 &42.5 $\pm$5.5 &10.2 &Yes&No\\
3&00$^h$58$^m$31.5$^s$&-72$^\circ$15$'$12$''$&31.6 $\pm$1.5 &21.5 &191.1 $\pm$24.5 &22.9 &No&No\\
4&00$^h$56$^m$42.0$^s$&-72$^\circ$23$'$29$''$&24.4 $\pm$1.2 &20.2 &142.0 $\pm$18.2 &20.0 &Yes&No\\
5&00$^h$56$^m$46.1$^s$&-72$^\circ$21$'$05$''$&21.9 $\pm$1.2 &18.0 &27.1 $\pm$3.6 &9.2 &Yes&No\\
6&00$^h$58$^m$38.8$^s$&-72$^\circ$35$'$00$''$&21.5 $\pm$2.2 &9.6 &10.4 $\pm$2.6 &5.8 &No&Yes\\
7&00$^h$58$^m$41.1$^s$&-72$^\circ$27$'$18$''$&21.3 $\pm$1.2 &17.6 &19.3 $\pm$2.6 &7.6 &Yes&No\\
8&00$^h$56$^m$54.3$^s$&-72$^\circ$16$'$35$''$&21.2 $\pm$1.7 &12.5 &156.6 $\pm$20.1 &22.6 &Yes&Yes\\
9&00$^h$58$^m$17.3$^s$&-72$^\circ$28$'$18$''$&18.6 $\pm$1.2 &15.0 &11.6 $\pm$1.9 &6.1 &No&No\\
10&00$^h$55$^m$59.9$^s$&-72$^\circ$21$'$34$''$&18.3 $\pm$1.3 &13.8 &12.3 $\pm$2.0 &6.2 &No&No\\
11&00$^h$58$^m$58.1$^s$&-72$^\circ$21$'$59$''$&17.4 $\pm$1.2 &14.3 &20.1 $\pm$2.7 &8.3 &No&No\\
12&00$^h$56$^m$12.9$^s$&-72$^\circ$23$'$46$''$&17.1 $\pm$1.2 &13.9 &12.7 $\pm$2.0 &6.5 &No&No\\
13&00$^h$58$^m$17.3$^s$&-72$^\circ$30$'$18$''$&16.9 $\pm$1.3 &13.4 &16.7 $\pm$2.3 &7.6 &No&No\\
14&00$^h$56$^m$49.5$^s$&-72$^\circ$30$'$23$''$&16.6 $\pm$1.3 &13.1 &14.4 $\pm$2.1 &7.1 &No&No\\
15&00$^h$56$^m$07.1$^s$&-72$^\circ$28$'$22$''$&16.6 $\pm$1.3 &12.4 &7.6 $\pm$1.8 &5.0 &Yes&No\\
16&00$^h$56$^m$03.6$^s$&-72$^\circ$23$'$46$''$&16.0 $\pm$1.2 &12.8 &10.2 $\pm$1.8 &5.9 &No&No\\
17&00$^h$56$^m$05.2$^s$&-72$^\circ$20$'$52$''$&15.9 $\pm$1.3 &11.9 &$<$4.9 $\pm$2.0 &4.1 &No&No\\
18&00$^h$56$^m$26.1$^s$&-72$^\circ$23$'$29$''$&15.8 $\pm$1.2 &13.0 &7.7 $\pm$1.7 &5.1 &Yes&No\\
19&00$^h$56$^m$00.1$^s$&-72$^\circ$19$'$58$''$&15.4 $\pm$1.4 &10.8 &14.8 $\pm$2.2 &7.2 &No&No\\
20&00$^h$58$^m$07.9$^s$&-72$^\circ$23$'$06$''$&15.0 $\pm$1.2 &12.6 &$<$5.0 $\pm$1.8 &4.2 &No&No\\
\enddata
\tablecomments{The columns give (1) source ID, (2) right ascension, (3) declination, (4) observed peak flux and noise level at 1.1 mm, (5) signal-to-noise ratio, (6) 1.1 mm total flux, (7) source radius, (8) whether the object associates with YSOs or $24~\micron$ objects, and (9) whether the object is located at the map edge or not.}
\end{deluxetable*}

\begin{deluxetable*}{rccccccccc}
\tabletypesize{\scriptsize}
\rotate
\tablewidth{22cm}
\tablecaption{Total flux densities of the 1.1 mm extended objects. \label{table:sed}}
\tablehead{Object ID&$S_\mathrm{1.1mm}$&$S_\mathrm{{\it Herschel} 500\mu m}$&$S_\mathrm{{\it Herschel} 350\mu m}$&$S_\mathrm{{\it Herschel} 250\mu m}$&$S_\mathrm{{\it Herschel} 160\mu m}$&$S_\mathrm{{\it Herschel} 100\mu m}$&$S_\mathrm{{\it Spitzer} 70\mu m}$&$S_\mathrm{{\it Spitzer} 24\mu m}$\\
&(mJy)&(mJy)&(mJy)&(mJy)&(mJy)&(mJy)&(mJy)&(mJy)}
\startdata
NEdeep-1 &  28.77  $\pm$ 4.05  &  219.57  $\pm$ 29.21  & 499.99  $\pm$ 42.50  &  865.16  $\pm$ 74.42  &  1348.15  $\pm$ 209.14  & $<$ 880.38  $+$ 294.55  & $<$ 3429.12  $+$ 172.15  & $<$ 117.67  $+$ 5.06  \\
2 &  42.50  $\pm$ 5.49  &  458.14  $\pm$ 58.91  & 1133.80  $\pm$ 145.45  &  2396.13  $\pm$ 307.31  &  5457.49  $\pm$ 781.40  &  7256.19  $\pm$ 1049.08  & $<$ 5666.17  $+$ 283.57  & $<$ 584.42  $+$ 23.42  \\
3 &  191.15  $\pm$ 24.48  &  1290.85  $\pm$ 165.33  & 2804.53  $\pm$ 224.40  &  4315.19  $\pm$ 345.31  &  5355.56  $\pm$ 538.37  &  1929.04  $\pm$ 216.38  & $<$ 14249.72  $+$ 712.51  & $<$ 393.60  $+$ 15.76  \\
4 &  142.01  $\pm$ 18.19  &  1233.59  $\pm$ 158.00  & 2746.50  $\pm$ 219.76  &  5144.05  $\pm$ 411.61  &  8448.07  $\pm$ 847.12  &  6929.23  $\pm$ 701.90  & $<$ 7295.19  $+$ 364.81  & $<$ 186.96  $+$ 7.52  \\
5 &  27.14  $\pm$ 3.56  &  313.51  $\pm$ 40.57  & 735.53  $\pm$ 59.61  &  1442.02  $\pm$ 116.86  &  2785.09  $\pm$ 310.21  &  1846.36  $\pm$ 306.06  & $<$ 1739.19  $+$ 88.00  & $<$ 59.43  $+$ 2.88  \\
6 &  10.37  $\pm$ 2.60  &  76.10  $\pm$ 14.50  & 172.45  $\pm$ 23.76  &  326.48  $\pm$ 45.17  & $<$ 564.61  $+$ 226.32  & $<$ 503.26  $+$ 389.14  & $<$ 595.94  $+$ 36.61  & $<$ 12.50  $+$ 2.61  \\
7 &  19.29  $\pm$ 2.64  &  244.03  $\pm$ 32.04  & 635.78  $\pm$ 52.16  &  1433.03  $\pm$ 116.86  &  3179.00  $\pm$ 358.57  &  4056.86  $\pm$ 502.43  & $<$ 3113.91  $+$ 156.56  & $<$ 1232.81  $+$ 49.35  \\
8 &  156.64  $\pm$ 20.06  &  1205.48  $\pm$ 154.40  & 2654.83  $\pm$ 212.43  &  4927.65  $\pm$ 394.30  &  8909.95  $\pm$ 892.75  &  7685.05  $\pm$ 774.90  & $<$ 9161.29  $+$ 458.10  & $<$ 223.78  $+$ 8.98  \\
9 &  11.58  $\pm$ 1.90  &  141.30  $\pm$ 20.15  & 314.92  $\pm$ 29.08  &  569.66  $\pm$ 53.67  &  948.30  $\pm$ 227.46  & $<$ 650.74  $+$ 375.03  & $<$ 635.30  $+$ 37.76  & $<$ 19.94  $+$ 2.58  \\
10 &  12.29  $\pm$ 2.00  &  96.39  $\pm$ 15.13  & 185.58  $\pm$ 20.72  &  316.58  $\pm$ 37.88  & $<$ 480.18  $+$ 207.40  & $<$ 597.94  $+$ 364.99  & $<$ 695.84  $+$ 40.08  & $<$ 14.18  $+$ 2.46  \\
11 &  20.06  $\pm$ 2.71  &  180.47  $\pm$ 24.00  & 361.14  $\pm$ 30.77  &  580.18  $\pm$ 50.81  &  750.48  $\pm$ 168.82  & $<$ 577.87  $+$ 276.31  & $<$ 923.12  $+$ 48.51  & $<$ 17.41  $+$ 1.93  \\
12 &  12.67  $\pm$ 1.96  &  81.42  $\pm$ 13.34  & 192.31  $\pm$ 20.55  &  355.29  $\pm$ 38.94  &  766.82  $\pm$ 208.93  & $<$ 517.02  $+$ 351.03  & $<$ 734.78  $+$ 41.45  & $<$ 15.37  $+$ 2.39  \\
13 &  16.67  $\pm$ 2.34  &  193.16  $\pm$ 25.74  & 466.69  $\pm$ 39.13  &  893.76  $\pm$ 75.06  &  1612.17  $\pm$ 231.38  &  1064.76  $\pm$ 314.97  & $<$ 884.74  $+$ 47.17  & $<$ 26.68  $+$ 2.24  \\
14 &  14.44  $\pm$ 2.12  &  133.41  $\pm$ 18.70  & 284.63  $\pm$ 25.97  &  524.53  $\pm$ 48.53  &  910.94  $\pm$ 198.90  & $<$ 835.20  $+$ 326.61  & $<$ 903.96  $+$ 48.45  & $<$ 19.09  $+$ 2.23  \\
15 &  7.59  $\pm$ 1.83  &  62.01  $\pm$ 13.45  & 128.06  $\pm$ 20.69  &  227.14  $\pm$ 39.44  & $<$ 321.88  $+$ 252.35  & $<$ 296.14  $+$ 447.59  & $<$ 341.51  $+$ 30.01  & $<$ 32.78  $+$ 3.24  \\
16 &  10.19  $\pm$ 1.79  &  76.26  $\pm$ 13.36  & 166.25  $\pm$ 20.01  &  302.98  $\pm$ 37.95  & $<$ 600.05  $+$ 220.76  & $<$ 360.79  $+$ 381.19  & $<$ 589.28  $+$ 36.17  & $<$ 11.29  $+$ 2.56  \\
17 & $<$ 4.91  $+$ 2.02  & $<$ 31.78  $+$ 14.03  & 73.52  $\pm$ 22.99  & $<$ 133.06  $+$ 44.58  & $<$ 246.95  $+$ 310.52  & $<$ 265.75  $+$ 552.96  & $<$ 324.77  $+$ 34.57  & $<$ 6.95  $+$ 3.68  \\
18 &  7.68  $\pm$ 1.69  &  54.02  $\pm$ 12.56  & 127.18  $\pm$ 19.93  &  232.85  $\pm$ 38.32  & $<$ 503.19  $+$ 250.32  & $<$ 446.56  $+$ 440.42  & $<$ 466.29  $+$ 33.62  & $<$ 18.65  $+$ 3.00  \\
19 &  14.80  $\pm$ 2.22  &  160.25  $\pm$ 21.92  & 379.77  $\pm$ 32.98  &  741.74  $\pm$ 64.37  &  1578.41  $\pm$ 236.04  &  1417.27  $\pm$ 343.46  & $<$ 1262.21  $+$ 65.43  & $<$ 33.48  $+$ 2.47  \\
20 & $<$ 5.02  $+$ 1.79  &  67.04  $\pm$ 15.42  & 150.84  $\pm$ 24.14  &  284.33  $\pm$ 46.78  & $<$ 579.83  $+$ 305.80  & $<$ 482.72  $+$ 538.76  & $<$ 405.30  $+$ 35.92  & $<$ 7.53  $+$ 3.58  \\
\enddata
\tablecomments{Two values provided in each cell show the representative fluxes and $1\sigma$ errors. We shows the upper limit values for the combination of the uniform and normal distribution in the MCMC analysis as the sign of "$<$".}
\end{deluxetable*}

\section{SED analysis}
\label{sec4}
As shown in Figure~\ref{fig_maps}, the 1.1~mm emission shows a reasonable spatial correlation to the \textit{Herschel} $100$--$500~\micron$ emission.
This indicates that these bands are emitted from the cold dust component that dominates the total dust mass.
In order to estimate the dust temperature and dust mass of the identified 1.1~mm objects, SED analysis was performed using the 1.1~mm, \textit{Herschel} 100, 160, 250, 350, and 500~$\micron$ data with the assumption of a single dust temperature.
We also used the photometry of the \textit{Spitzer} $24~\micron$ and $70~\micron$ as upper limits.
The total flux density of cold thermal dust emission at a wavelength $\lambda$, $S_{\mathrm{obs},\lambda}$, can be modeled by \begin{equation}
S_{\mathrm{model},\lambda} = \kappa_{\mathrm{dust}, \lambda} B_\lambda(T_{\mathrm{dust}}) M_{\mathrm{dust}} D^{-2},
\end{equation}
where $\kappa_{\mathrm{dust}, \lambda}$ is the emissivity of dust grains, $B_\lambda$ is the Planck function, $M_{\mathrm{dust}}$ is the total dust mass, $T_{\mathrm{dust}}$ is the dust temperature, and $D=60$~kpc is the distance to the SMC \citep[e.g.,][]{2005MNRAS.357..304H}.
For the emissivity of the cold dust component, we used $\kappa_{\mathrm{dust}, \lambda}=12.5\times (160~\micron/\lambda)^{\beta}~\mathrm{cm^2~g^{-1}}$ \citep{2007ApJ...657..810D,2014ApJ...780..172D}.
We estimated the posterior distributions of $T_\mathrm{dust}$, $M_\mathrm{dust}$, and the index of emissivity $\beta$ using Markov chain Monte Carlo (MCMC) method.
The likelihood function is defined as 
\begin{equation}
L=\prod_{\lambda} L_\lambda,
\end{equation}
and
\begin{equation}
L_\lambda=\exp(-\frac{(S_{\mathrm{obs},\lambda}-S_{\mathrm{mod},\lambda})^2}{2\sigma_{\mathrm{obs},\lambda}^2}).
\end{equation}
For the upper limits of photometry, we used the one-sided Gaussian distribution:
\begin{equation}
L_\lambda \propto \left\{
    \begin{aligned}
        &1&    (0\leq S_{\mathrm{model},\lambda} \leq S_{\lambda,obs}) \\
        &\exp(-\frac{(S_{\mathrm{obs},\lambda}-S_{\mathrm{mod},\lambda})^2}{2\sigma_{\mathrm{obs},\lambda}^2})&(S_{\mathrm{model},\lambda} > S_{\mathrm{obs},\lambda})
   \end{aligned}
   \right
..
\end{equation}
We assumed the uniform prior probability distributions for the fitting parameters, with the ranges of $0<T_\mathrm{dust} (\mathrm{K})<60$, $0<\log(M_\mathrm{dust}/M_\odot)<4$, and $0<\beta<5$. 
We used \texttt{PyMC3} \citep{salvatier2016probabilistic} to implement the MCMC method.

By just selecting a previously known gas-to-dust ratio $GDR=1000$ \citep[with a possible error of a factor of 2,][]{2007ApJ...658.1027L,2011A&A...536A..17P,2014ApJ...797...85G,2014ApJ...797...86R}, the total gas masses $M_\mathrm{gas}$ were estimated by
\begin{equation}
M_\mathrm{gas}=GDR\times M_\mathrm{dust}.
\end{equation}

\begin{figure}
\epsscale{1.2}
\plotone{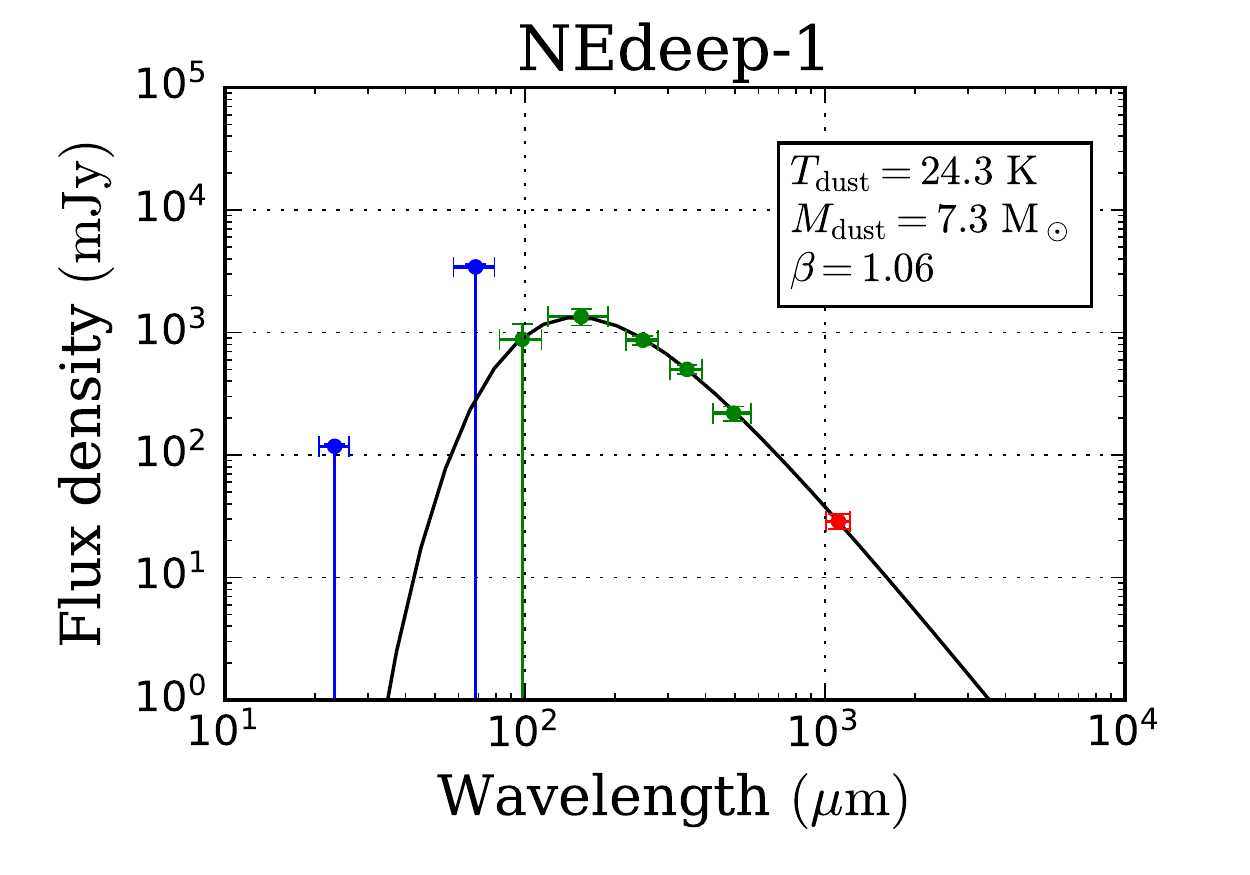}
\caption{SEDs of AzTEC/SMC NEdeep objects. The solid line represents the maximum likelihood SED models for the cold dust component. The red (1.1~mm), green (\textit{Herschel} 100, 160, 250, 350, and 500~$\micron$), and blue (\textit{Spitzer} 24 and 70~$\micron$) points represent the fitting points for the cold dust SED. The fitting parameters of the maximum likelihood model are shown in each figures. (The complete figure set is shown in the appendix.)\label{fig_sed}}
\end{figure}

SEDs of the 1.1~mm objects are presented in Figure~\ref{fig_sed} and the obtained physical properties are summarized in Table~\ref{table:sedfit}.

\begin{deluxetable*}{rcccccc}
\tabletypesize{\scriptsize}
\tablewidth{12cm}
\tablecaption{Physical properties of the 1.1 mm extended objects. \label{table:sedfit}}
\tablehead{Object ID&$T_\mathrm{dust}$&$M_\mathrm{dust}$&$M_\mathrm{gas}$&$\beta$&$n_\mathrm{H_2}$&$N_\mathrm{H_2}$\\
&(K)&($\mathrm{M_\sun}$)&($\times 10^3 \mathrm{M_\sun}$)& &($\mathrm{H_2/cm^3}$)&($\mathrm{\times 10^{20} H_2/cm^2}$)}
\startdata
NEdeep-1&$24.3\pm2.3$&$<8.6$&$<8.6$&$1.06\pm0.14$&$<180.0$&$<59.2$\\
2&$33.0\pm2.3$&$10.7^{+2.8}_{-2.2}$&$10.7^{+2.8}_{-2.2}$&$1.24\pm0.11$&$35.7^{+9.3}_{-7.4}$&$15.0^{+3.9}_{-3.1}$\\
3&$19.5\pm0.7$&$70.4^{+14.1}_{-11.7}$&$70.4^{+14.1}_{-11.7}$&$1.12\pm0.11$&$21.0^{+4.2}_{-3.5}$&$19.8^{+4.0}_{-3.3}$\\
4&$26.3\pm1.3$&$35.2^{+7.2}_{-6.0}$&$35.2^{+7.2}_{-6.0}$&$1.09\pm0.10$&$15.7^{+3.2}_{-2.7}$&$12.9^{+2.6}_{-2.2}$\\
5&$23.4\pm1.2$&$16.3^{+3.7}_{-3.0}$&$16.3^{+3.7}_{-3.0}$&$1.47\pm0.11$&$75.1^{+17.2}_{-14.0}$&$28.4^{+6.5}_{-5.3}$\\
6&$25.2\pm5.5$&$<9.7$&$<9.7$&$1.12\pm0.31$&$<529.1$&$<126.8$\\
7&$30.1\pm1.8$&$8.4^{+1.9}_{-1.5}$&$8.4^{+1.9}_{-1.5}$&$1.46\pm0.10$&$69.6^{+15.6}_{-12.7}$&$21.7^{+4.9}_{-4.0}$\\
8&$29.1\pm1.6$&$25.1^{+5.3}_{-4.4}$&$25.1^{+5.3}_{-4.4}$&$0.93\pm0.10$&$7.8^{+1.7}_{-1.4}$&$7.2^{+1.5}_{-1.3}$\\
9&$20.7\pm2.7$&$<18.1$&$<18.1$&$1.54\pm0.19$&$<868.5$&$<216.9$\\
10&$23.6\pm5.1$&$<9.3$&$<9.3$&$1.00\pm0.24$&$<411.1$&$<105.3$\\
11&$20.1\pm2.5$&$<16.2$&$<16.2$&$1.22\pm0.18$&$<303.8$&$<103.7$\\
12&$30.8\pm3.6$&$<1.8$&$<1.8$&$0.86\pm0.13$&$<72.2$&$<19.2$\\
13&$22.8\pm1.6$&$11.0^{+3.3}_{-2.6}$&$11.0^{+3.3}_{-2.6}$&$1.49\pm0.13$&$91.1^{+27.7}_{-21.2}$&$28.4^{+8.6}_{-6.6}$\\
14&$26.0\pm3.7$&$<6.5$&$<6.5$&$1.12\pm0.17$&$<195.8$&$<57.2$\\
15&$22.3\pm4.6$&$<8.6$&$<8.6$&$1.17\pm0.27$&$<732.6$&$<151.3$\\
16&$25.8\pm4.6$&$<5.1$&$<5.1$&$1.01\pm0.20$&$<266.5$&$<64.8$\\
17&$17.6\pm4.5$&$<21.9$&$<21.9$&$1.88\pm0.57$&$<3516.3$&$<587.4$\\
18&$24.2\pm4.5$&$<6.0$&$<6.0$&$1.12\pm0.25$&$<477.4$&$<100.5$\\
19&$27.9\pm2.3$&$5.0^{+1.5}_{-1.2}$&$5.0^{+1.5}_{-1.2}$&$1.28\pm0.12$&$48.8^{+14.8}_{-11.3}$&$14.4^{+4.4}_{-3.3}$\\
20&$18.9\pm5.3$&$<55.2$&$<55.2$&$1.90\pm0.50$&$<8122.2$&$<1396.1$\\\tableline
\enddata
\tablecomments{The columns give (1) Source ID, (2) dust temperature, (3) total dust mass, (4) total gas mass, (5) index of emissivity (6) $\mathrm{H_2}$ density, and (7) $\mathrm{H_2}$ column density. The errors and upper limits are provided by 1$\sigma$ and 3$\sigma$, respectively. }
\end{deluxetable*}

In Section \ref{sec6}, we also attempted an image-based SED fit to discuss the detailed structures of the detected clumps.
Using the same posterior distributions of dust temperature and index of dust emissivity, and the uniform posterior for the column density of cold dust $N_\mathrm{dust}$ in the range of $-10<\log(N_\mathrm{dust}/(\mathrm{g~cm^{-2}}))<-3$, we estimate the parameters using the following equation: 
\begin{equation}
I_\lambda = \kappa_{\mathrm{dust}, \lambda} B_\lambda(T_{\mathrm{dust}}) N_{\mathrm{dust}},
\end{equation}
where $I_\lambda$ is the flux density of each pixel.

We should pay attention to the possibility of larger $\kappa_{\mathrm{dust}, \lambda}$. In the SMC and LMC, \citet{2017ApJ...837...98G} suggested that $\kappa_{\mathrm{dust},160\micron}=30.2~\mathrm{cm^2~g^{-1}}$, which is about three times larger than that of some physically motivated models \citep[e.g.,][]{2007ApJ...657..810D}. This causes a dust mass estimate about three times lower than our $\kappa_{\mathrm{dust},\lambda}$ value.

\section{Physical properties of NEdeep objects}
\label{sec5}
\subsection{Comparison with the wide-survey objects}
The 1.1~mm objects identified by \citet[][hereafter wide-survey objects]{2017ApJ...835...55T} are potential candidates of dense gas clouds that are forming or about to form massive stars.
In the wide-survey objects, only NE-1 was detected as a counterpart of the NEdeep-1 object in the NE field.
The sensitivity of the wide-survey is about $6~\mathrm{mJy~beam^{-1}}$ in this region, which is not sufficient to detect the other NEdeep objects with $>5\sigma$.
Thus, the high-sensitivity NEdeep observations provide information about many objects that are undetectable in the wide-survey.

Table~\ref{table4} presents the typical physical properties of the NEdeep and wide-survey objects. We did not include the three objects in the NEdeep samples located at the edge of the 1.1~mm images, and N88-1 of the wide-survey objects in the statistics.
\citet{2017ApJ...835...55T} assumed a fixed $\beta=1.2$ for estimating the dust temperature and mass of the wide-survey objects.
Here, we estimated the average and standard deviation values of $\beta=1.3\pm0.3$ for the NEdeep objects, and this difference between free $\beta$ and $\beta=1.2$ fits causes the systematic average differences of 5\% and 8\% for dust temperature and mass (except for upper limit values) estimate, respectively.
Thus, roughly speaking, sets of physical properties between the wide-survey and NEdeep objects are able to be compared directly.

Comparing the physical properties between the wide-survey and NEdeep objects, statistical physical properties of the NEdeep objects, other than the mass and size, are roughly consistent with those of the wide-survey objects. 
In contrast, the mass and size of the NEdeep objects are relatively smaller than wide-survey objects, which indicates that the high-sensitivity observation makes it possible to detect smaller mass objects. 
These low-mass objects are also good candidates for dense gas clumps that are connected to massive star formation, because about 40\% of the NEdeep objects host YSOs, including lower mass objects, such as NEdeep-15 and 18.

The dust temperature of the NEdeep objects may be slightly lower than that of the wide-survey objects.
The difference of star forming activity between the NEdeep and wide-survey objects would explain this difference, because the wide-survey objects include very active star forming regions such as Lirs49,
SMCB-2N, and N66, which show the dust temperature of $\sim 40$~K \citep{2017ApJ...835...55T}. 
We avoid further discussion regarding the comparison with the wide-survey objects here, because of the difference of the SED fit models.

The observed area in the SMC NE region is only about 2\% of that of the wide field, and therefore, a more sensitive ($\sim 1~\mathrm{mJy~rms}$) and wider (a few square degrees) survey of the SMC at 1.1~mm will provide several hundred samples of dust-selected clouds, down to a gas mass of $1\times 10^3~\mathrm{M_\odot}$.
It is also important for high-resolution observations to detect gas clumps of $< 10^3~\mathrm{M_\odot}$, because the sizes of many NEdeep objects are very close to the diffraction limit.

\begin{deluxetable}{lcc}
\tabletypesize{\scriptsize}
\tablewidth{11cm}
\tablecaption{Summary of physical properties between the SMC NEdeep and wide survey objects.\label{table4}}
\tablehead{&This study (NEdeep)&SMC wide survey}
\startdata
Cloud number&17&43\\
$T_\mathrm{dust}$ range&17.6--33.0~K&17--45~K\\
$T_\mathrm{dust}$ ave.$\pm$std.&24.3$\pm$4.4~K&28.7$\pm$4.4~K\\
$\beta$ range&0.9--1.9&1.2 (fixed) \\
$\beta$ ave.$\pm$std.&1.29$\pm$0.29&--\\
$M_\mathrm{gas}$ range&(5.0--70)$\times 10^3~\mathrm{M_\sun}$&(4.1--336)$\times 10^3~\mathrm{M_\sun}$\\
$M_\mathrm{gas}$ median&$11.0\times 10^3~\mathrm{M_\sun}$&$44.6\times 10^3~\mathrm{M_\sun}$\\
$R$ range &4--23~pc&6--40~pc\\
$R$ median&7.1~pc&11.8~pc\\
$\mathrm{H_2}$ density range&16--91~$\mathrm{cm^{-3}}$&17--171~$\mathrm{cm^{-3}}$\\
$\mathrm{H_2}$ density ave.$\pm$std. dev.&$49\pm 29~\mathrm{cm^{-3}}$&$68\pm 36~\mathrm{cm^{-3}}$\\
$\mathrm{H_2}$ column density range&(13--28)$\times 10^{20}~\mathrm{cm^{-2}}$&(10--44)$\times 10^{20}~\mathrm{cm^{-2}}$\\
$\mathrm{H_2}$ column density ave.$\pm$std. dev.&(20$\pm$6)$\times 10^{20}~\mathrm{cm^{-2}}$&(29$\pm$8)$\times 10^{20}~\mathrm{cm^{-2}}$\\
\enddata
\tablecomments{Upper limit values of $M_\mathrm{gas}$, $\mathrm{H_2}$ density, and $\mathrm{H_2}$ column density are excluded from the statistics.}
\end{deluxetable}

\subsection{Index of emissivity and $\beta-T_\mathrm{dust}$ relation}
Here, we investigate the characteristics of the index of emissivity in detail.
Firstly, as shown in Table~\ref{table:sedfit} and \ref{table4}, the range of $\beta$ is 0.9--1.9. This is consistent with the reasonable range of single temperature dust particles, $1<\beta<2$, motivated by the Kramers-Kronig relation at long wavelengths and the emissivity model for silicates \citep[e.g.,][]{1984ApJ...285...89D}.

Secondly, previous studies, using lower resolution datasets, show $\beta\sim 1.2$ in the SMC \citep{2003ApJ...596..273A,2007ApJ...658.1027L,2010A&A...523A..20B,2010A&A...519A..67I,2011A&A...536A..17P}.
The obtained average ($\pm$ standard deviation) value $\beta_\mathrm{ave.}=1.3\pm 0.3$ in the NEdeep objects is consistent with these studies.
On the other hand, the wide range of $\beta$ values among the NEdeep objects suggests that the existence of a complex temperature structure or difference of dust composition affects the index of emissivity of each dust-selected cloud.

\begin{figure}
\epsscale{1.3}
\plotone{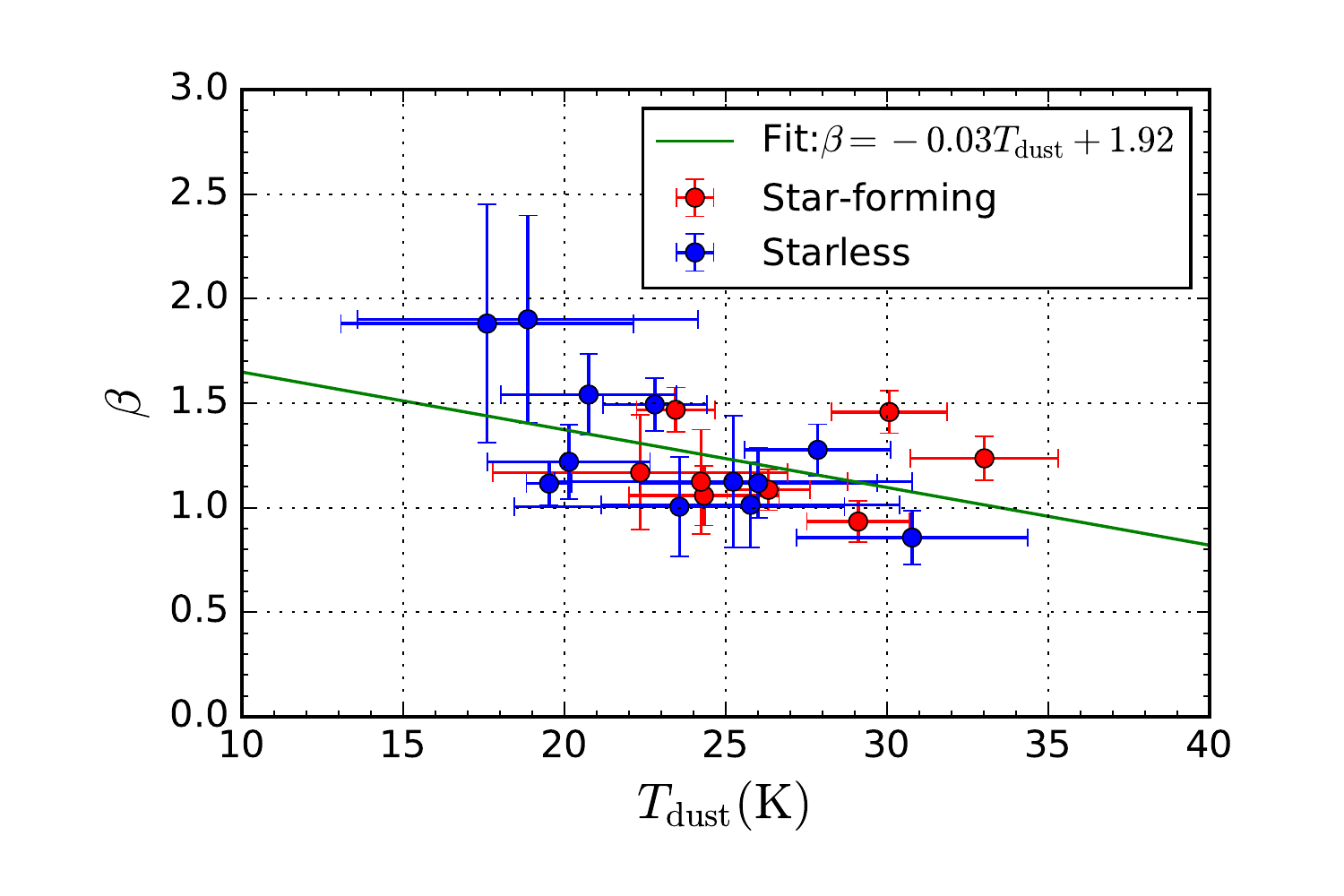}
\caption{The relation between dust temperature and index of emissivity.\label{fig_Tbeta}} \label{Tbeta}
\end{figure}

Finally, the relation between dust temperature and index of emissivity is shown in Figure~\ref{fig_Tbeta}.
As a result of fitting with a linear function, we obtained the relationship:
\begin{equation}
\beta(T_\mathrm{dust})=(-0.03\pm0.01)T_\mathrm{dust}+(1.93\pm0.34).
\end{equation}
The negative correlation of the $\beta-T_\mathrm{dust}$ relation has already reported by \citet{{2014ApJ...797...85G}} in the SMC, using the pixel-based SED fit with the minimum $\chi^2$ method.
Some studies revealed that the minimum $\chi^2$ method is very susceptible to data noise, and makes the negative $\beta-T_\mathrm{dust}$ relation \citep[e.g.,][]{2009ApJ...696..676S,2009ApJ...696.2234S}. 
On the other hand, the MCMC method can significantly reduce the noise-induced $\beta-T_\mathrm{dust}$ correlations \citep{2012ApJ...752...55K,2013A&A...556A..63J}. 
Thus, our result suggests that the negative $\beta-T_\mathrm{dust}$ relation in the NEdeep objects is intrinsic, and its origin can be attributed to the difference in physical structures or dust properties.

Some possible interpretations of the negative $\beta-T_\mathrm{dust}$ relation have been proposed by model and simulation studies of molecular cores \citep[e.g.,][]{2009ApJ...696..676S,2011A&A...530A.101M,2012A&A...539A..71J}. 
The difference of line-of-sight temperature structures of dust-selected clouds, particularly internally-heated objects, creates a negative $\beta-T_\mathrm{dust}$ relation.
Although this picture is consistent with the dust temperature and \textit{Spitzer} $24~\micron$ relation, which supports the fact that the heating source of the dust-selected clouds is mainly local star formation activity \citep{2017ApJ...835...55T}, we cannot find a clear relation between existence of star formation activity in the $\beta-T_\mathrm{dust}$ relation, as shown in Figure~\ref{Tbeta}.
On the other hand, the photometric errors and band selections also cause a weak negative or positive bias, even in the case of MCMC fit.
Further investigation using analytic or simulation modeling, and larger and more sensitive observation of dust-selected clouds, are necessary to reveal origin of the $\beta-T_\mathrm{dust}$ relation.

\section{Comparison with CO emission and star formation tracer}
\label{sec6}
The 1.1~mm objects already detected by the NANTEN and Mopra CO observations in the observation region exhibit a weaker star formation activity than the other CO-selected molecular clouds in the SMC. Therefore, these objects may remain at the initial state of molecular cloud evolution without the influence of star formation. Thus, these 1.1~mm objects are very important to understand the relationship between the chemical and dynamical evolution of molecular clouds under the low-metallicity environment.

In this section, we reveal the characteristics of the molecular clouds that have already reported CO detection \citep{2001PASJ...53L..45M, 2010ApJ...712.1248M} by comparing the gas mass obtained by the CO data and distribution of YSOs.
In addition, we investigate the internal structure of these objects by comparing the CO and PAH distributions and the result of the map-based SED analysis of dust continuum data.

\subsection{Gas mass estimate from CO}
\label{sec61}

The gas masses estimated from the SED fit of the thermal dust continuum are not taken into consideration for the possible bias caused by the gas-to-dust ratio, emissivity, and temperature distributions in the objects.
Therefore, it is important to compare the gas masses estimated by the CO luminosity as another tracer of gas mass to check the consistency.
We estimated the gas masses of the NEdeep-4, 5, 7, 9, and 13 objects using the Mopra CO data assuming an $X_\mathrm{CO}$ factor by the following steps.
First, we estimated the CO luminosity within the contours of the 1.1~mm objects and velocity range of the CO line.
Second, we estimated the gas masses using the following equation:
\begin{eqnarray}
M_\mathrm{gas, CO} (\mathrm{M_\sun}) &=& 2~\mu m_\mathrm{p} X_\mathrm{CO} L_\mathrm{CO} \\
                                           &=& 21.8~L_\mathrm{CO} (\mathrm{K~km~s^{-1}~pc^2}), 
\end{eqnarray}
where the $X_\mathrm{CO}$ factor is $1\times 10^{21}~\mathrm{cm^{-2}~(K~km~s^{-1})^{-1}}$, which is reported by \citet{2010ApJ...712.1248M} in the northeast region.
This $X_\mathrm{CO}$ factor is also consistent with the estimate from the NANTEN CO and \textit{Herschel} dust continuum \citep{2001PASJ...53L..45M,2014ApJ...797...86R}.  

\begin{figure}
\epsscale{1.2}
\plotone{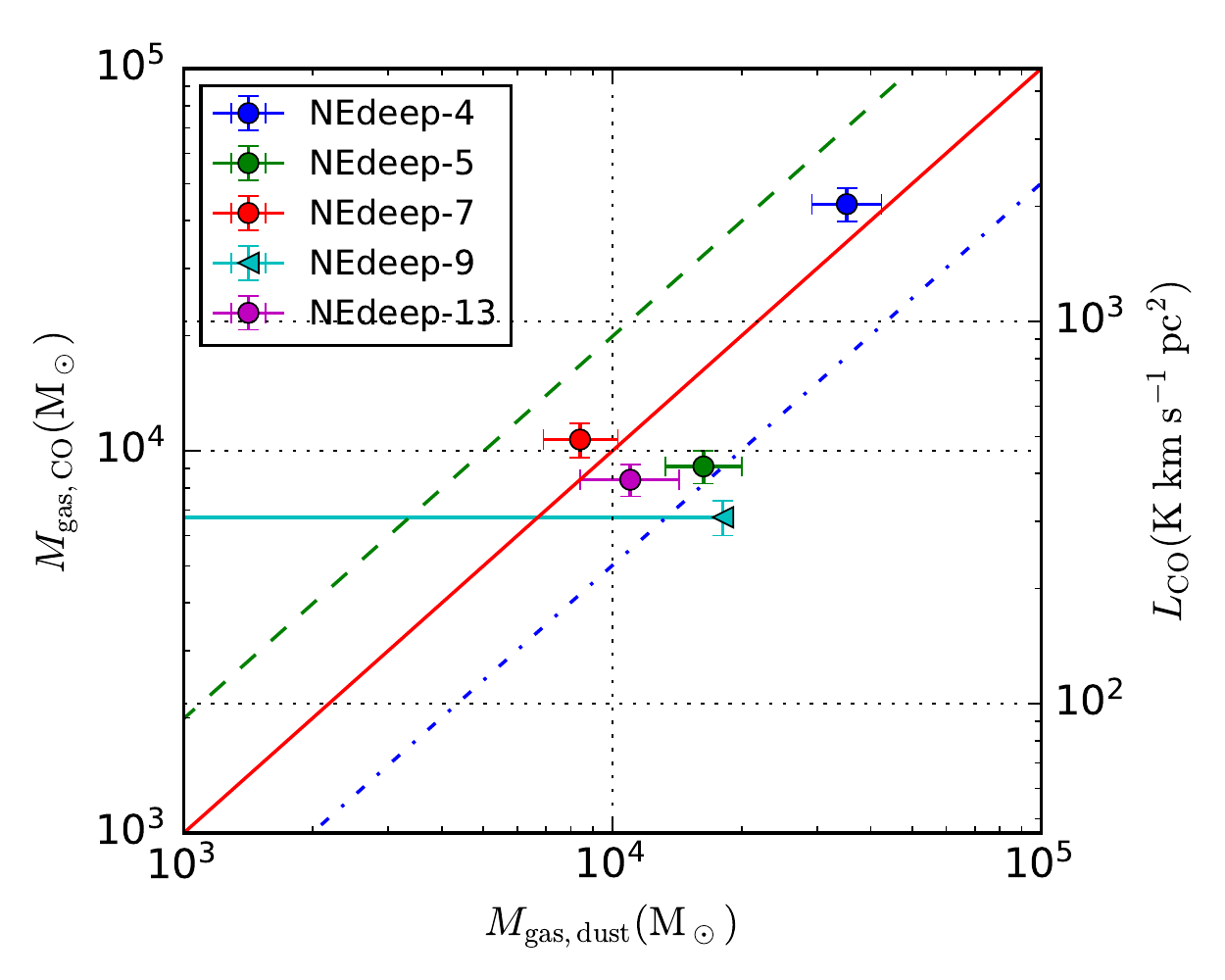}
\caption{Comparison of gas mass estimated by dust and CO ($X_\mathrm{CO}=1\times 10^{21}~\mathrm{cm^{-2}~(K~km~s^{-1})^{-1}}$) of NEdeep objects. The range of CO luminosity is also shown on the right vertical axis. The lines show $M_\mathrm{gas,CO}=aM_\mathrm{gas,dust}$ with $a=1$, 2, and 0.5 in solid (red), dashed (green), and dashed-dotted lines (blue), respectively. \label{fig_mass}}
\end{figure}

A summary of gas mass estimates by CO and dust is presented in Table~\ref{table5} and Figure~\ref{fig_mass}.
The estimated gas masses by CO and dust are consistent with each other within the errors of a factor of 2.
Therefore, the gas masses estimated from the dust continuum can be used to estimate the total CO luminosity of dense clouds in the SMC.
Thus, the assumption of a gas-to-dust ratio of 1000, and the $X_\mathrm{CO}$ factor of $1\times 10^{21}~\mathrm{cm^{-2}~(K~km~s^{-1})^{-1}}$, with uncertainties of the factor of 2, respectively, are reliable to estimate CO or continuum fluxes for future GMC studies.
In contrast, these samples are biased in CO-detected objects, and the dust-selected clouds that have not been detected by CO lines are not included.
Therefore, it is important to conduct further CO line observations on the CO-dark dust-selected cloud samples reported in this study and \citet{2017ApJ...835...55T}.

\begin{deluxetable}{lcccc}
\tabletypesize{\scriptsize}
\tablewidth{12cm}
\tablecaption{Comparison of gas masses estimated from dust and CO.\label{table5}}
\tablehead{Object ID&$M_\mathrm{gas, dust}$&$M_\mathrm{gas, CO}$&$M_\mathrm{gas, dust}/M_\mathrm{gas, CO}$&$L_\mathrm{CO}$\\
&($10^3~\mathrm{M_\odot}$)&($10^3~\mathrm{M_\odot}$)&&($10^2~\mathrm{K~km~s^{-1}~pc^2}$)}
\startdata
NEdeep-4&$35.2^{+7.2}_{-6.0}$&$44.2\pm4.4$&0.80&$20.3\pm2.3$\\
NEdeep-5&$16.3^{+3.7}_{-3.0}$&$9.1\pm0.9$&1.79&$4.2\pm0.4$\\
NEdeep-7&$8.4^{+1.9}_{-1.5}$&$10.7\pm1.1$&0.79&$4.9\pm0.5$\\
NEdeep-9&$<18.1$&$6.7\pm0.7$&--&$3.1\pm0.3$\\
NEdeep-13&$11.0^{+3.3}_{-2.6}$&$8.4\pm0.8$&1.31&$3.9\pm0.4$\\
\enddata
\end{deluxetable}

\subsection{Internal structure of the 1.1~mm objects}
\label{sec62}

\begin{figure*}
\epsscale{1.2}
\plotone{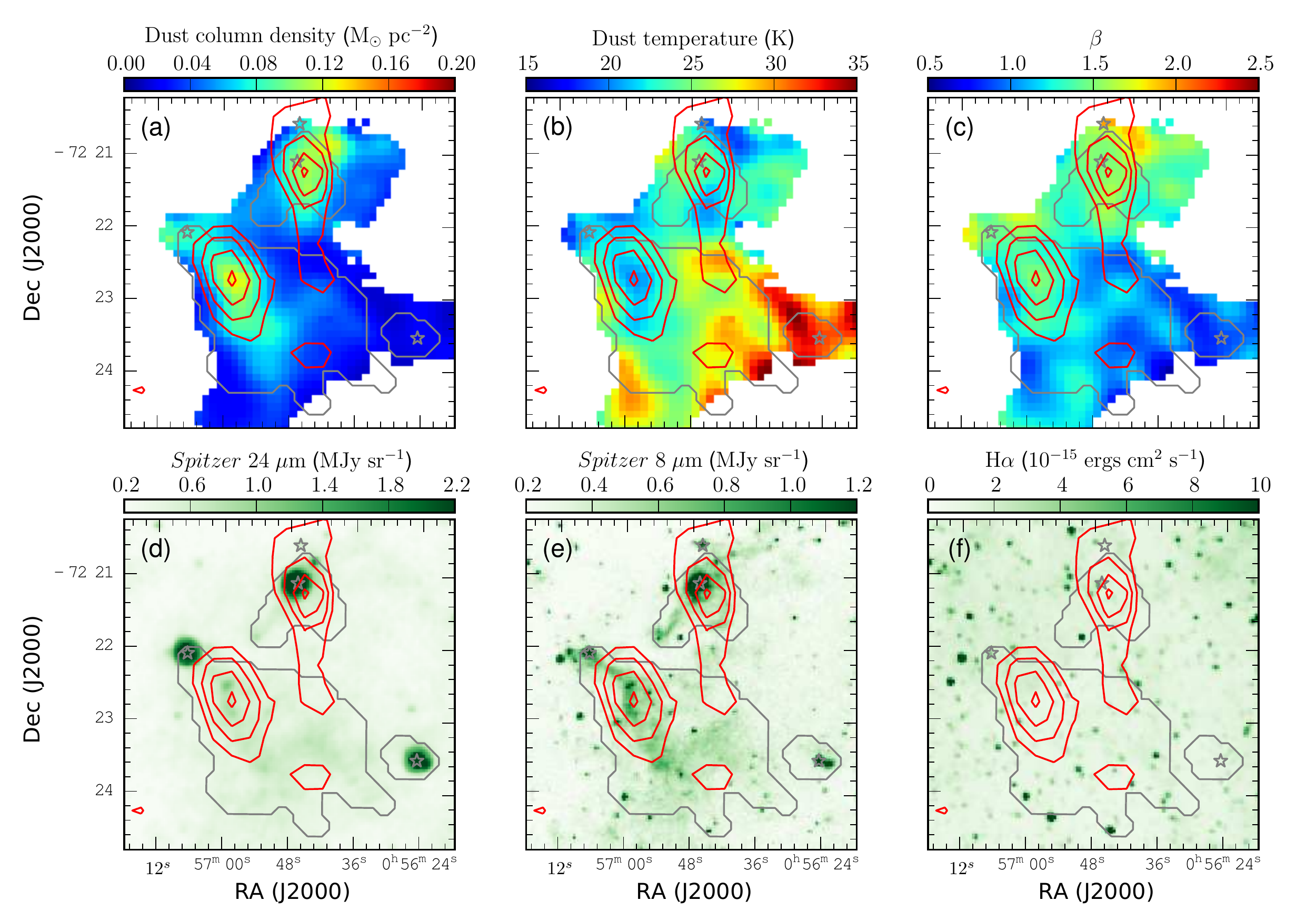}
\caption{Spatial distribution of (a) Dust column density, (b) dust temperature, (c) index of emissivity, (d) \textit{Spitzer} $24~\micron$, (e) \textit{Spitzer} $8~\micron$, and (f) $\mathrm{H\alpha}$ in the NEdeep-4, 5, and 11 regions. The Mopra CO emission is represented as red (150~$\mathrm{km~s^{-1}}$ component) contours with a step of 0.5~$\mathrm{K~km~s^{-1}}$ starting from 2.5~$\mathrm{K~km~s^{-1}}$. The gray contours represent the edges of the AzTEC 1.1~mm objects. The star symbols indicate the positions of YSOs \citep{2007ApJ...655..212B, 2013ApJ...778...15S}.
\label{NEdeep-4518}}
\end{figure*}

\begin{figure*}
\epsscale{1.2}
\plotone{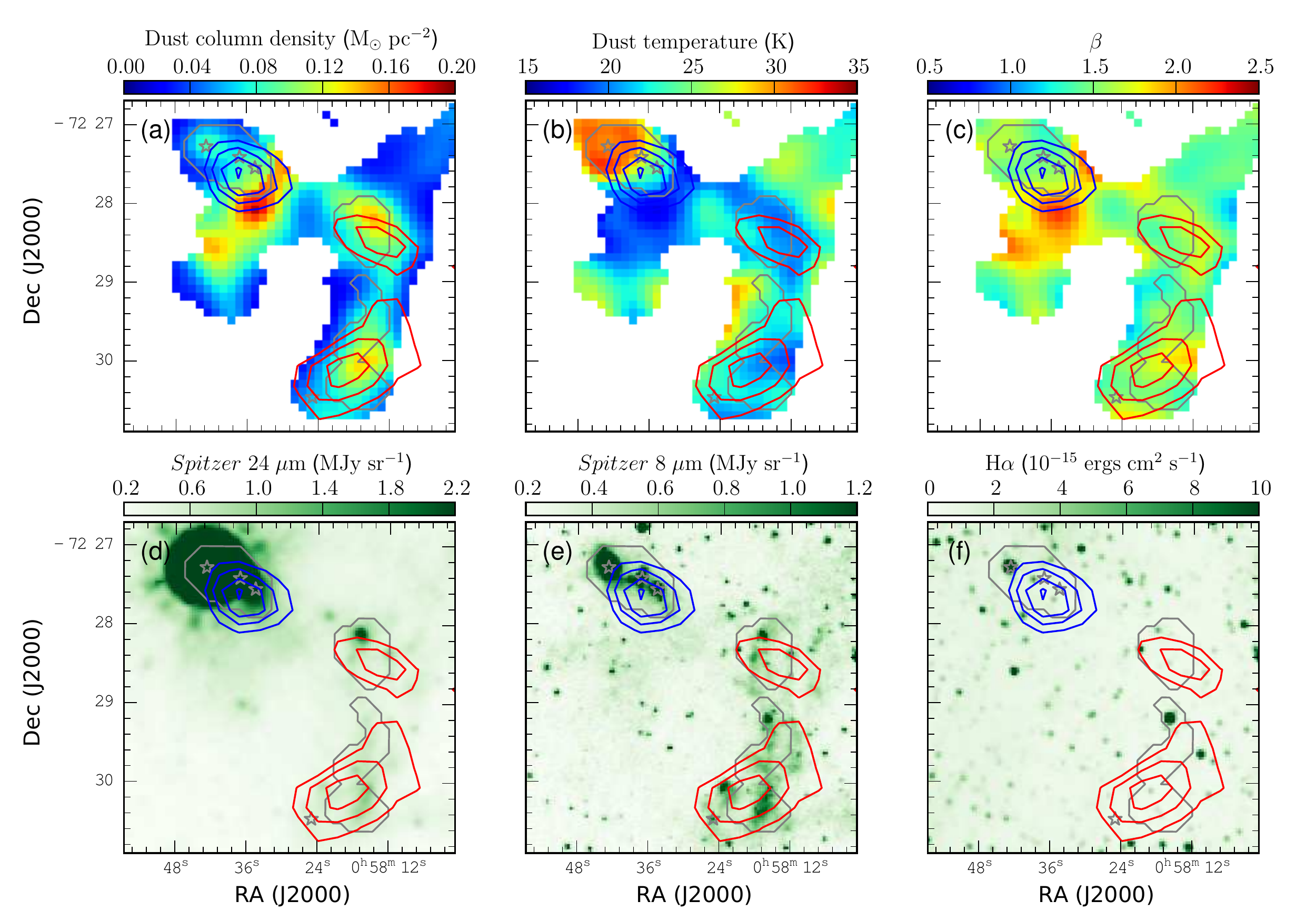}
\caption{Spatial distribution of (a) Dust column density, (b) dust temperature, (c) index of emissivity, (d) \textit{Spitzer} $24~\micron$, (e) \textit{Spitzer} $8~\micron$, and (f) H$\alpha$ in the NEdeep-7, 9, and 13 regions. The Mopra CO emission is represented as red (150~$\mathrm{km~s^{-1}}$ component) and blue (120~$\mathrm{km~s^{-1}}$ component) contours with a step of 0.5~$\mathrm{K~km~s^{-1}}$ starting from 2.5~$\mathrm{K~km~s^{-1}}$. The gray contours, star symbols are the same as the previous figure.
\label{NEdeep-7913}}
\end{figure*}

We focus on the relationship between the dust, CO, and star formation in five CO-detected dust-selected clouds in the SMC NE region by resolving the spatial structures.
Figures \ref{NEdeep-4518} and \ref{NEdeep-7913} present the distributions of dust properties (dust column density, dust temperature, and index of emissivity) and star formation tracers (\textit{Spitzer} $24~\micron$, $8~\micron$, and H$\alpha$) of the 1.1~mm objects that have already been detected by CO ($J=1$--$0$) by NANTEN and Mopra \citep{2001PASJ...53L..45M, 2010ApJ...712.1248M}.

\subsubsection{Star formation activity} 
\label{sec621}
Here, we will examine the star formation activities of the CO-detected dust-selected clouds.
First, extended H$\alpha$ emission is not observed in these dust clouds at a resolution of about 1~pc ($3''$--$4''$). In contrast, YSOs or bright $24~\micron$ objects are present inside or near the dust-selected clouds.
In particular, NEdeep-7 has a bright YSO at $24~\micron$ on the east side of the object, and dust temperature becomes high.
Except for this object, we cannot find radio continuum emission at 8.64 and 4.8~GHz.
In addition to the bright YSO in the NEdeep-7, a YSO at the peak of NEdeep-5 also may affects the ISM because the dust temperature around this YSO is slightly higher than that of the other high column density regions.
Although the existence of point-like H$\alpha$ objects cannot disprove the existence of very compact H{\small II} regions, we can say that NEdeep-4, 9, and 13 are young evolution phases of the GMCs that are not affected by strong UV radiation from massive young stars.
Therefore, these dust-selected clouds can be good candidates for investigation of the initial conditions for massive star and cluster formation under low-metallicity environments.

In the CO-detected dust-selected clouds in the SMC NE region, we cannot find reliable starless objects.
\citet{2017ApJ...835...55T} also reported the lack of CO-detected and starless objects in the dust-selected cloud samples selected in the full SMC.
A possible explanation is that the timescale of star formation (2--3~Myr) is shorter than that of CO molecule formation in the low-metallicity ISM, as pointed out by a numerical study of \citet{2012MNRAS.426..377G}.

\subsubsection{Coincidence of peak positions among dust, CO, and star formation} 
\label{sec622}

In Figures~\ref{NEdeep-4518} and \ref{NEdeep-7913}, we notice that the CO-detected objects, except for NEdeep-7, show good agreement with the distributions between CO and the dust column density estimated from dust.
This suggests that both CO and dust column density effectively trace dense molecular gas regions in GMCs.
On the other hand, the positions of YSOs do not correspond to CO and dust column peaks.
In particular, NEdeep-7 shows the distances between YSO and the dust peak of about 10~pc ($\sim 30''$), which are sufficiently larger than the pointing errors of $<5''$, $<3''$, and $<1''$ for Mopra CO, cold dust (ASTE 1.1~mm and \textit{Herschel}), and star formation tracers (\textit{Spitzer}, H$\alpha$), respectively.
This implies that the strong UV radiation from YSOs in NEdeep-7 affects CO and cold dust distribution, but should be investigated by high resolution studies in detail.

An interesting fact, particularly shown in NEdeep-4, is that the filamentary structure traced by $8~\micron$, which mainly traces emission from PAHs, shows good correlation with the CO emission. 
The spatial correlation of PAHs and CO emission has already been pointed out by previous low-metallicity ISM studies.
\citet{2010ApJ...715..701S} demonstrated that the PAH fraction spatially correlates with the CO intensity with a resolution of about 50~pc in the SMC.
Recently, a low-metallicity ($\sim 0.2~Z_\sun$) dwarf galaxy NGC~6822 was observed by a CO line using ALMA with a resolution of about 2~pc, reporting that the CO emission shows a better correlation with $8~\micron$ than $24~\micron$ and H$\alpha$ \citep{2017ApJ...835..278S}.
These studies support the notion that PAHs emissions effectively trace a photo-dissociated surface of dense molecular gas clouds observed by CO and dust continuum.
Our result also indicates that the PAH clumps or filamentary structures are good candidates of CO-emitting regions in the low-metallicity ISM, although it also should be confirmed by high resolution study using ALMA.

We should also note that the extended dust component is found outside the PAH structures or CO emission.
The most extended emission at 24 and 8~$\micron$ shows a good correspondence with the 1.1~mm objects.
We can understand that this extended dust emission traces the photo-dissociated surface of barely evolved gas clouds.
The high dust column density implies the ability to form massive stars in the future but may not have yet formed compact gravitationally bound filament/clumps.
In such regions, CO molecules would also not have formed yet, because of the long formation timescale of CO, as suggested by \citet{2012MNRAS.426..377G}.

\section{Summary}
\label{sec7}
The main results of this study are summarized below.
\begin{enumerate}
\item{We obtained a 1.1~mm image by the AzTEC instrument on the ASTE telescope toward the SMC NE regions with an effective resolution of $40''$ ($\sim12$~pc).
A median rms noise level of $1.3~\mathrm{mJy beam^{-1}}$ was achieved for a field of 343 square arcminutes ($\sim 20'\times20'$).}
\item{We identified 20 objects in the observation region. Two NANTEN CO clouds that were not detected in a previous 1.1~mm survey were detected and resolved into multiple dust-selected clouds.}

\item{The dust mass and temperature were estimated by SED analysis using the MCMC method with the 1.1~mm, \textit{Herschel}, and \textit{Spitzer} data.
Although the gas and dust masses of twelve 1.1~mm objects were estimated as upper limits, the other eight objects show the gas mass range of $5\times 10^3$--$7\times 10^4~\mathrm{M_\odot}$, assuming a gas-to-dust ratio of 1000.
The ranges of dust temperature and index of emissivity were 18--33~K and 0.9--1.9, respectively.}
\item{The 1.1~mm objects discovered by this study (NEdeep objects) show smaller dust masses and lower dust temperatures than the shallower 1.1~mm survey of \citet{2017ApJ...835...55T}.
The fact that 40\% of the 1.1~mm objects host YSOs, including relatively low-mass dust-selected clouds, suggests that the 1.1~mm objects trace dense gas clumps related to massive star formation.}
\item{The average of the index of emissivity is comparable to previous low resolution studies in the SMC. The relation between the dust temperature and the index of emissivity shows a slightly negative correlation.}
\item{We investigated five dust-selected clouds that have already been detected by CO in detail. The total gas masses of the 1.1~mm objects estimated from the Mopra CO data are comparable with the gas masses estimated by the SED analysis of thermal dust emission.
$X_\mathrm{CO}=1\times 10^{21}~\mathrm{cm^{-2} (K~km~s^{-1})^{-1}}$ and a gas-to-dust ratio of 1000, with uncertainties of a factor of 2, respectively, are reliable for the estimate of the total gas mass of molecular or dust-selected clouds each other in the SMC.}
\item{We compared the internal structure of dust-selected clouds estimated by an image-based SED fit with the Mopra CO and various star formation tracers. 
These objects exhibit no extended H$\alpha$ emission, although they were associated with YSOs or $24~\micron$ point sources, suggesting that these objects are young GMCs where star formation has just started; these are important targets to investigate the initial environment of massive star formation in low-metallicity environments.
Dust column density show good spatial correlation with CO emission except for NEdeep-7.
The $8~\micron$ filamentary structures and clumps show a similar spatial distribution with the CO emission and dust column density estimated by the image-base SED fits, implying that the filamentary structures or compact clumps traced by PAH emission are very good candidates of CO emitters.
The extended emission at 24 and 8$~\micron$, which do not show CO emission, exhibits a similar spatial distribution to the 1.1~mm objects, also suggesting that the cold gas component not yet affected by the gravitational contraction in GMCs are also traced by the emission from very small grains and PAH.}
\end{enumerate}

To obtain a more detailed understanding of the relation among cold dust, CO, and PAH emission, it is essential to conduct high-resolution CO and dust continuum observations toward the dust-selected clouds by ALMA, with a resolution comparable to the \textit{Spitzer}/IRAC bands ($\sim 1''$).

\acknowledgments
The ASTE project was driven by NRO/NAOJ, in collaboration with the University of Chile and Japanese institutes including the University of Tokyo, Nagoya University, Osaka Prefecture University, Ibaraki University, Hokkaido University, and Joetsu University of Education.
Observations with ASTE were carried out remotely from Japan using NTT's GEMnet2 and its partner R\&E networks, which are based on the AccessNova collaboration between the University of Chile, NTT Laboratories, and NAOJ.
This work is based on data products made with the \textit{Spitzer} Space Telescope (JPL/Caltech under a contract with NASA).
Data analysis was, in part, carried out on the open-use data analysis computer system at the Astronomy Data Center, ADC, of the National Astronomical Observatory of Japan.
This research made use of the SIMBAD database, operated at CDS, Strasbourg, France.
This research made use of Astropy, a community-developed core Python package for Astronomy \citep{2013A&A...558A..33A}.
This study was supported by the MEXT Grant-in-Aid for Specially Promoted Research JP20001003 and the JSPS Grant-in-Aid for Scientific Research (S) JP17H06130. M.R. acknowledges support from CONICYT (CHILE) through FONDECYT grant N$^\mathrm{o}$1140839.

\facilities{ASTE, \textit{Spitzer} (MIPS, IRAC), \textit{Herschel} (SPIRE, PACS), NANTEN, MOPRA, ATCA, Parkes}
\software{IDL, Astropy \citep{2013A&A...558A..33A}, PyMC3 \citep{salvatier2016probabilistic}, NumPy \citep{walt2011numpy}, SciPy \citep{scipy2001}, Matplotlib \citep{Hunter:2007}}
\bibliographystyle{yahapj}
\bibliography{SMCII}

\newpage

\appendix
\section{Spectral energy distribution of the 1.1~mm objects}
\label{appA}

\begin{figure}
\epsscale{1.1}
\plottwo{SMCNE_SED_1.pdf}{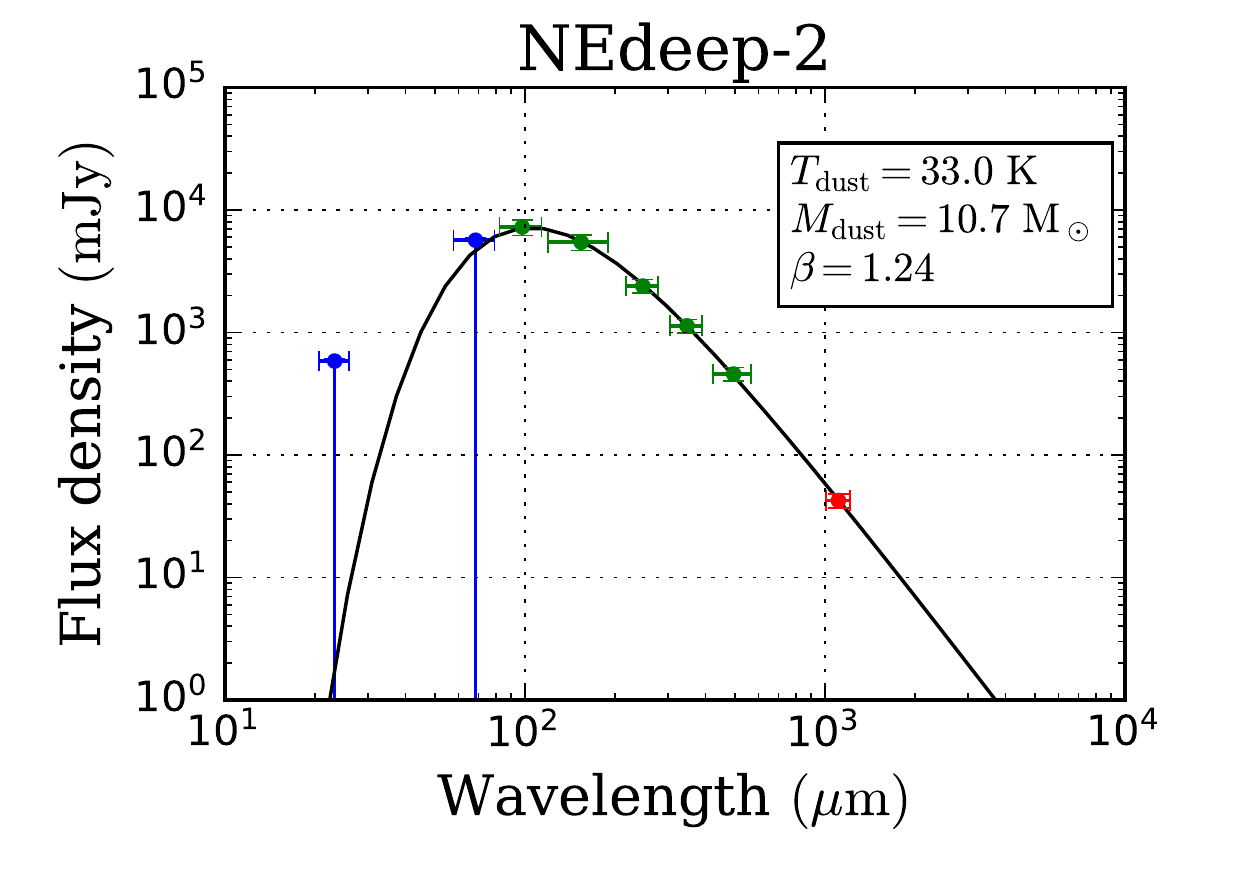}
\plottwo{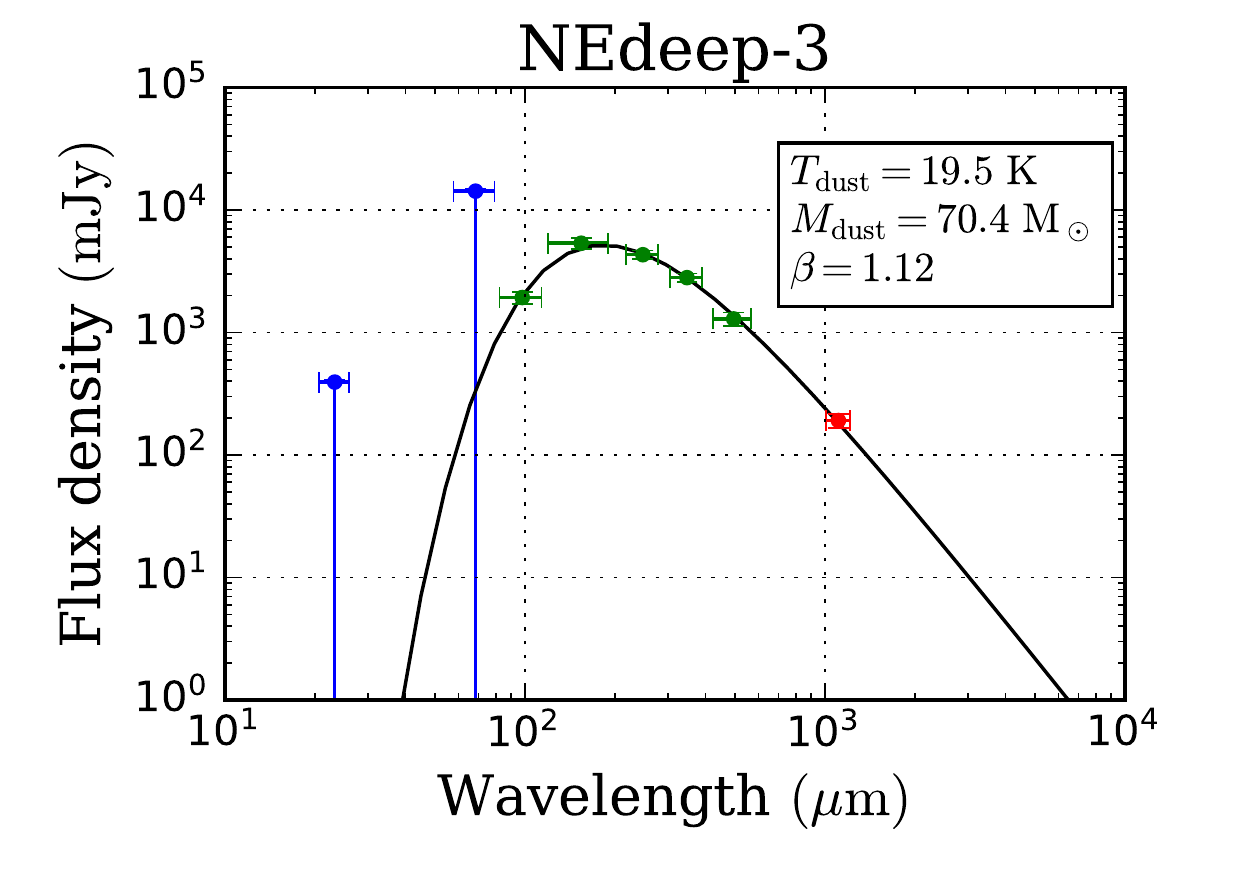}{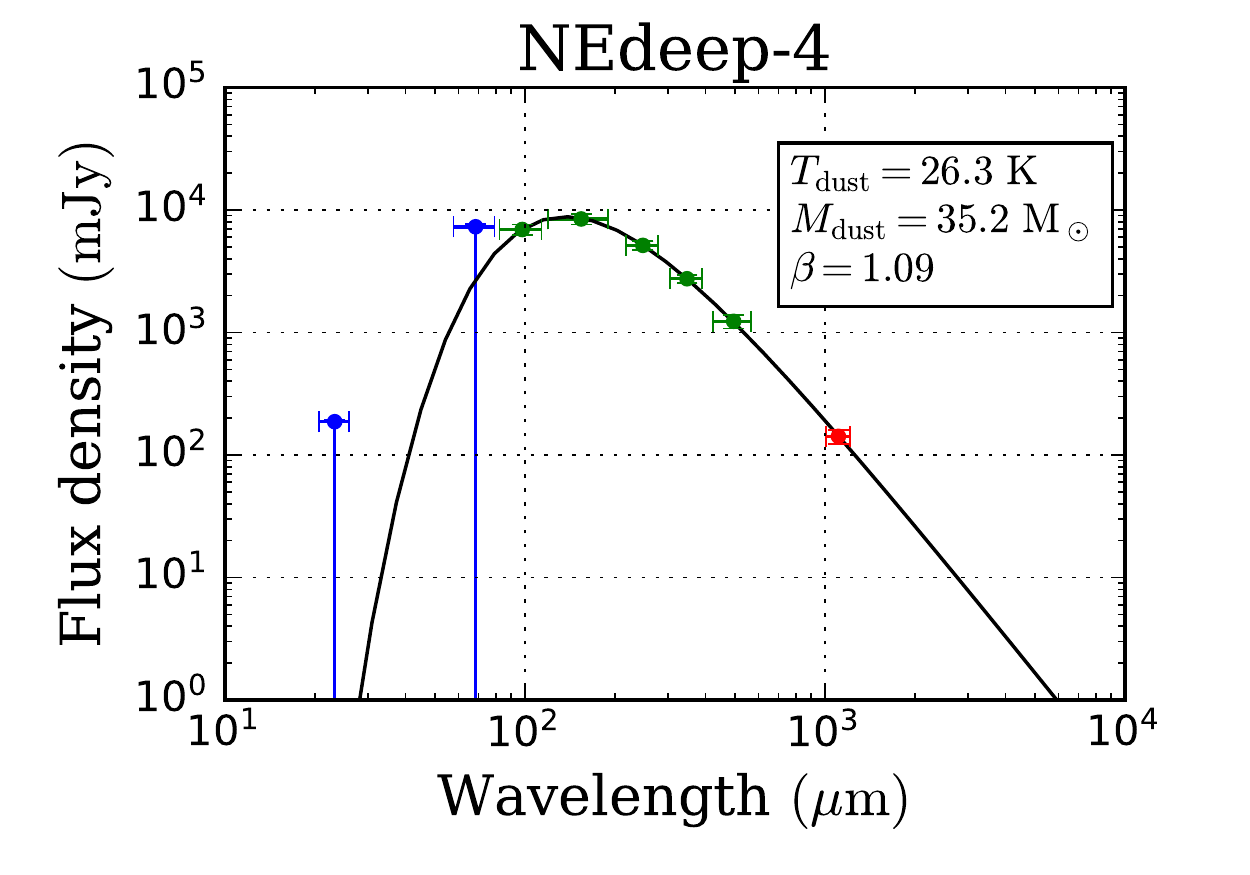}
\plottwo{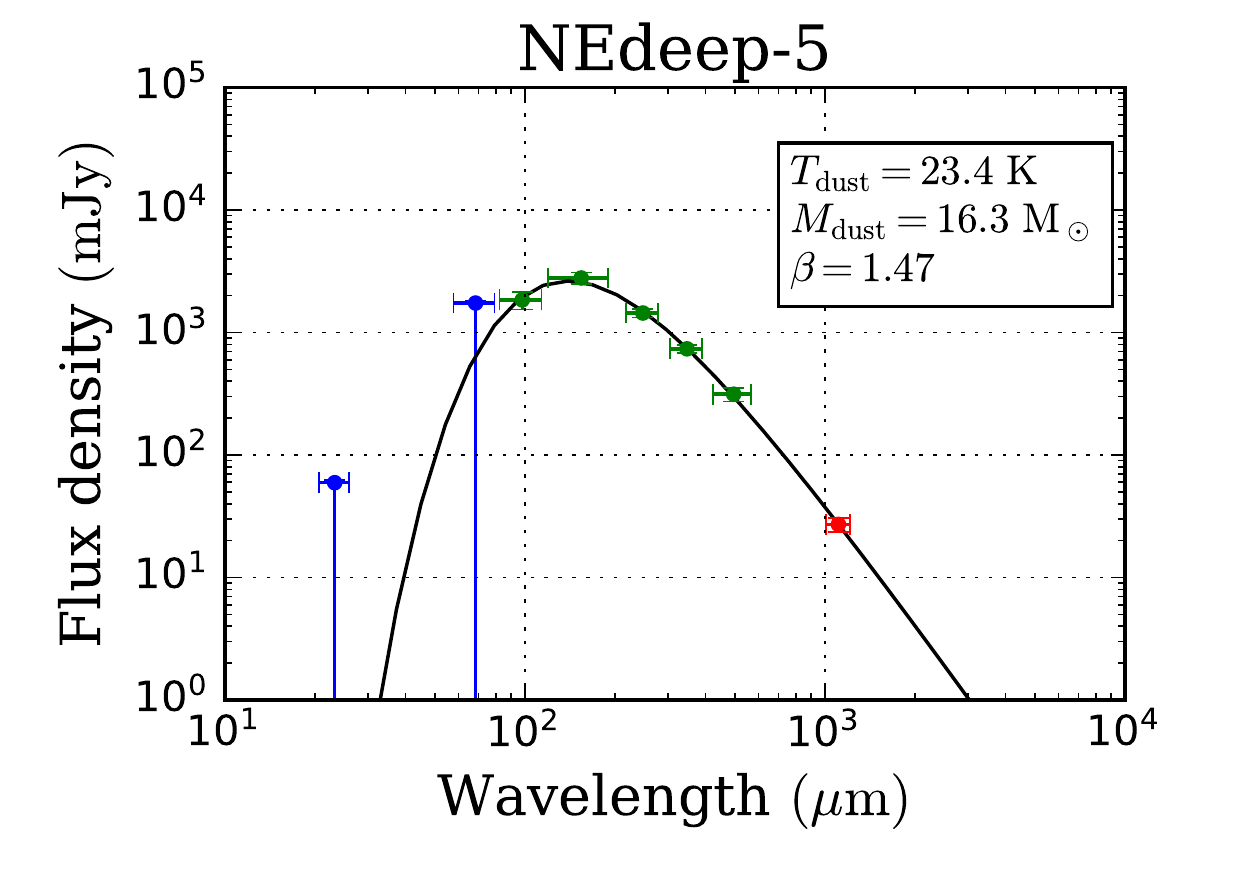}{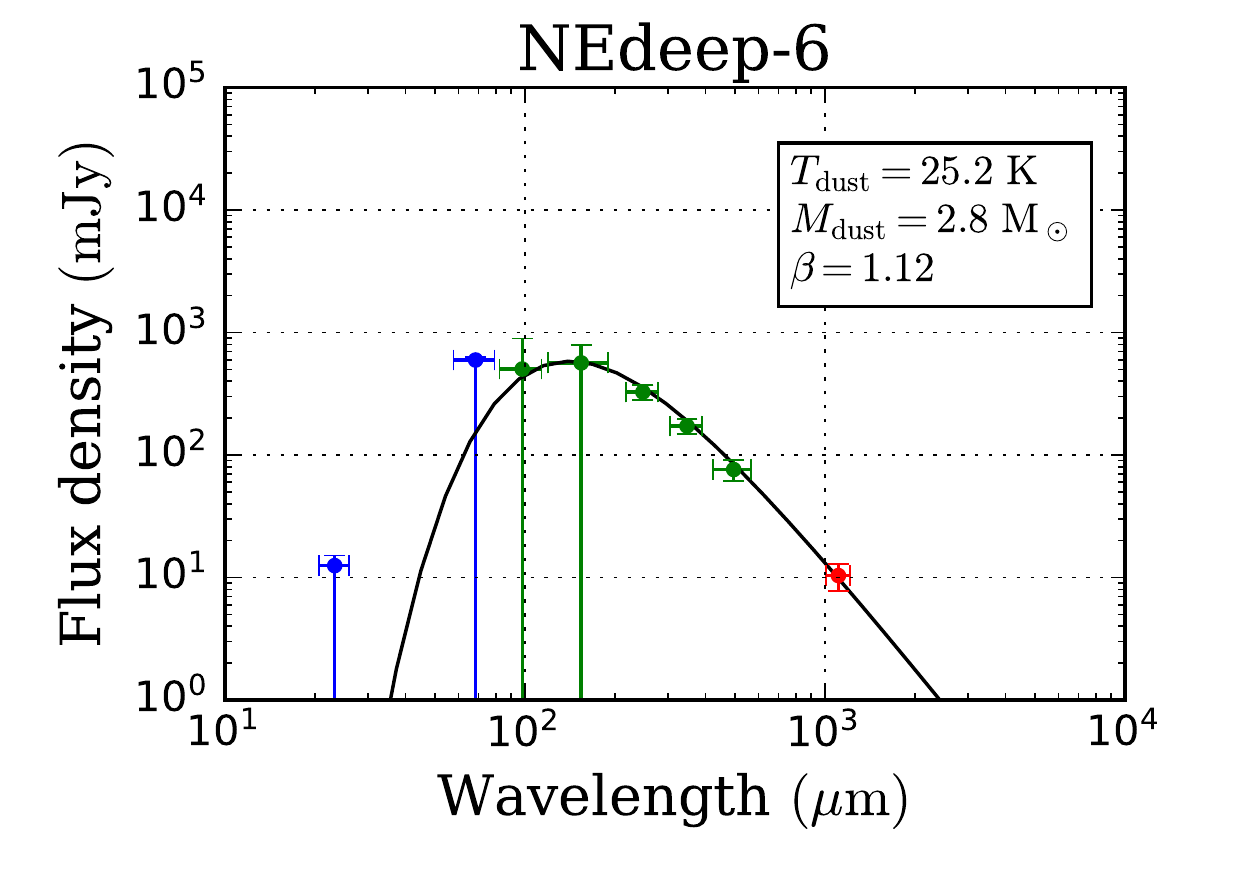}
\plottwo{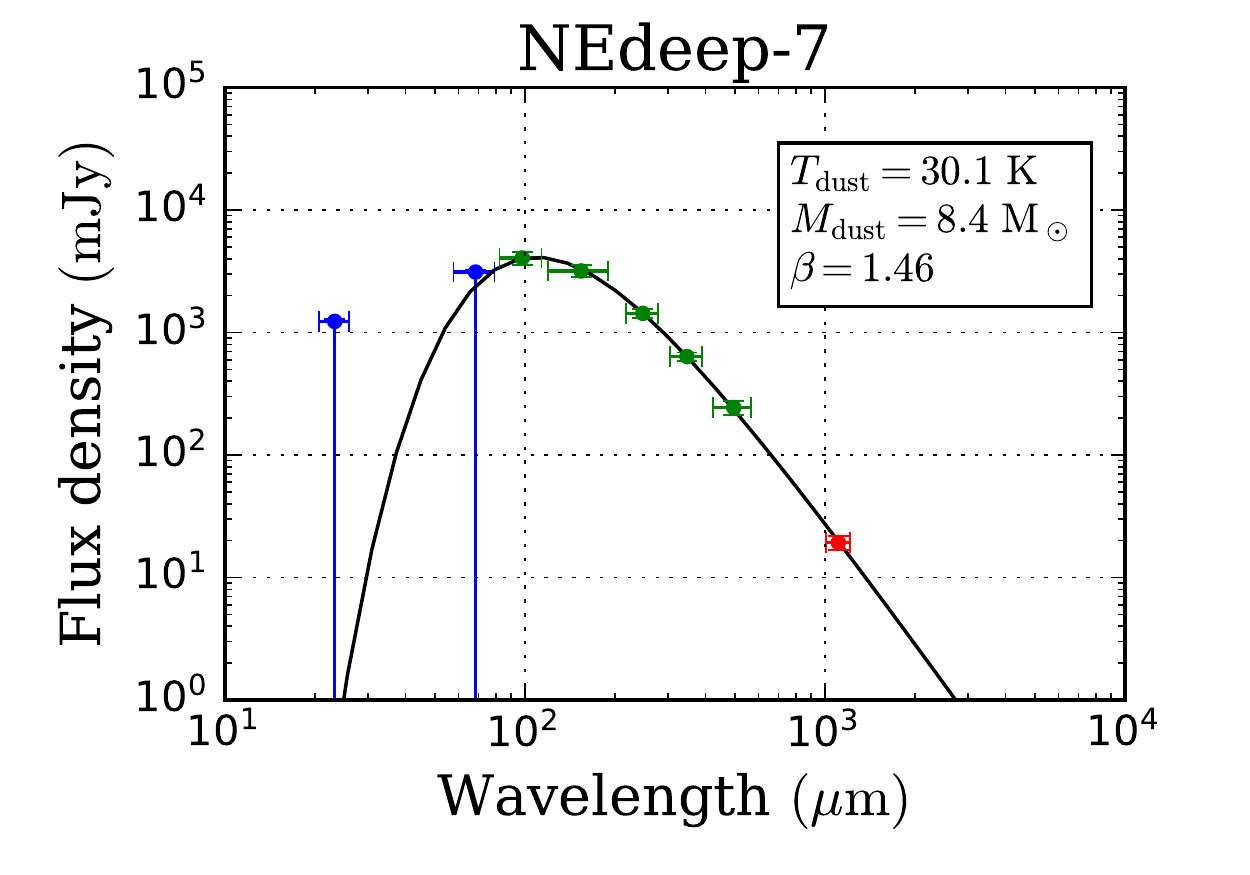}{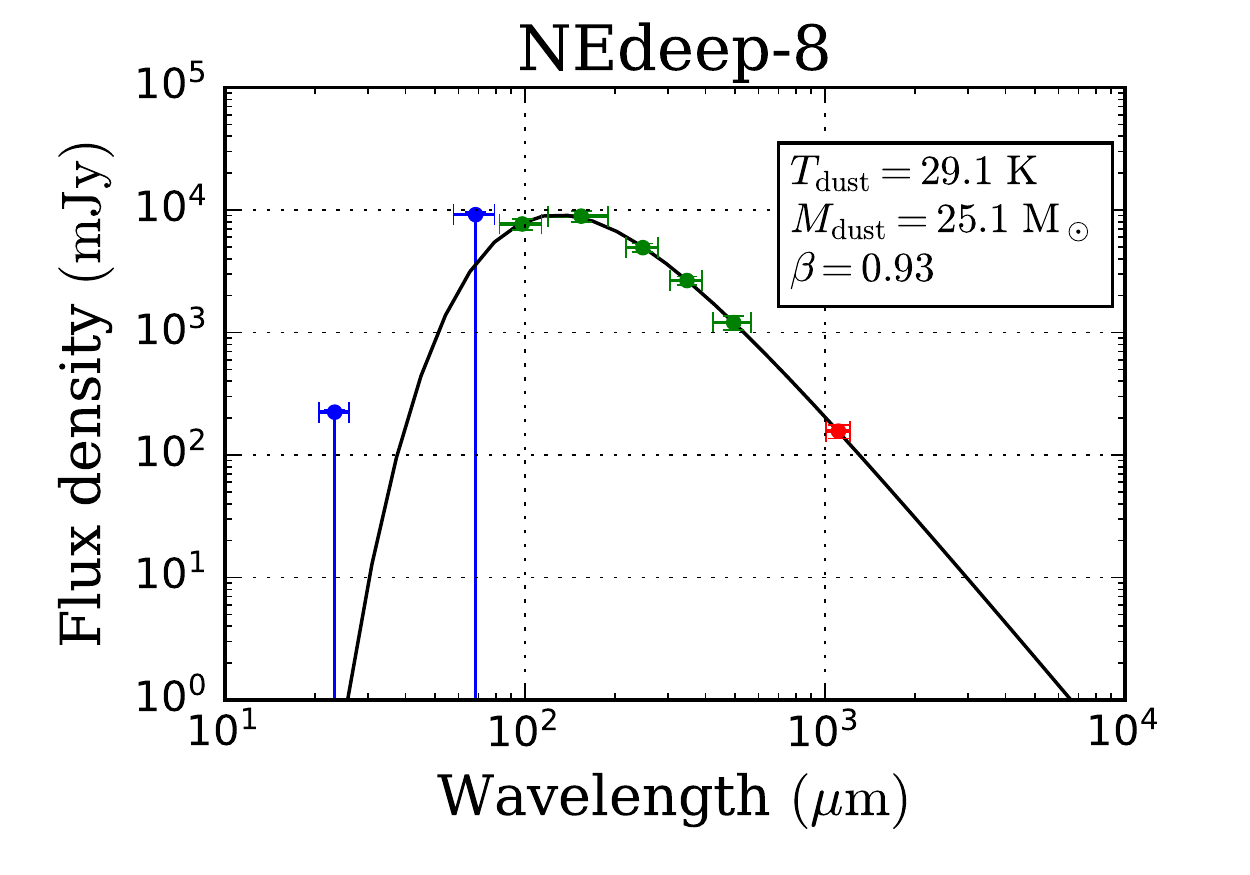}
\caption{SEDs of NEdeep objects. The solid line represents the maximum likelihood SED models for the cold dust component. The red (1.1~mm), green (\textit{Herschel} 100, 160, 250, 350, and 500~$\micron$), and blue (\textit{Spitzer} 24 and 70~$\micron$) points represent the fitting points for the cold dust SED. The fitting parameters of the maximum likelihood model are shown in each figure.}
\end{figure}

\begin{figure}
\epsscale{1.1}
\plottwo{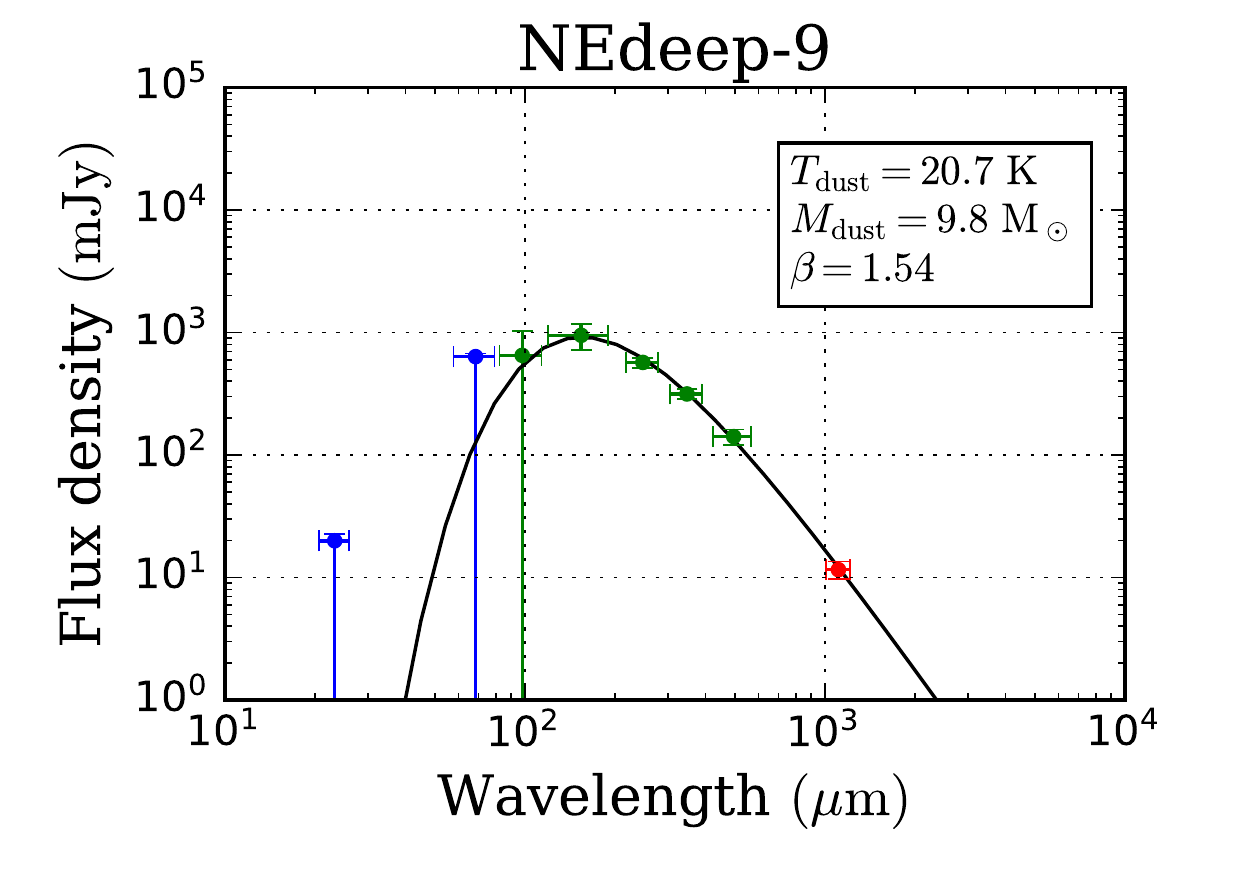}{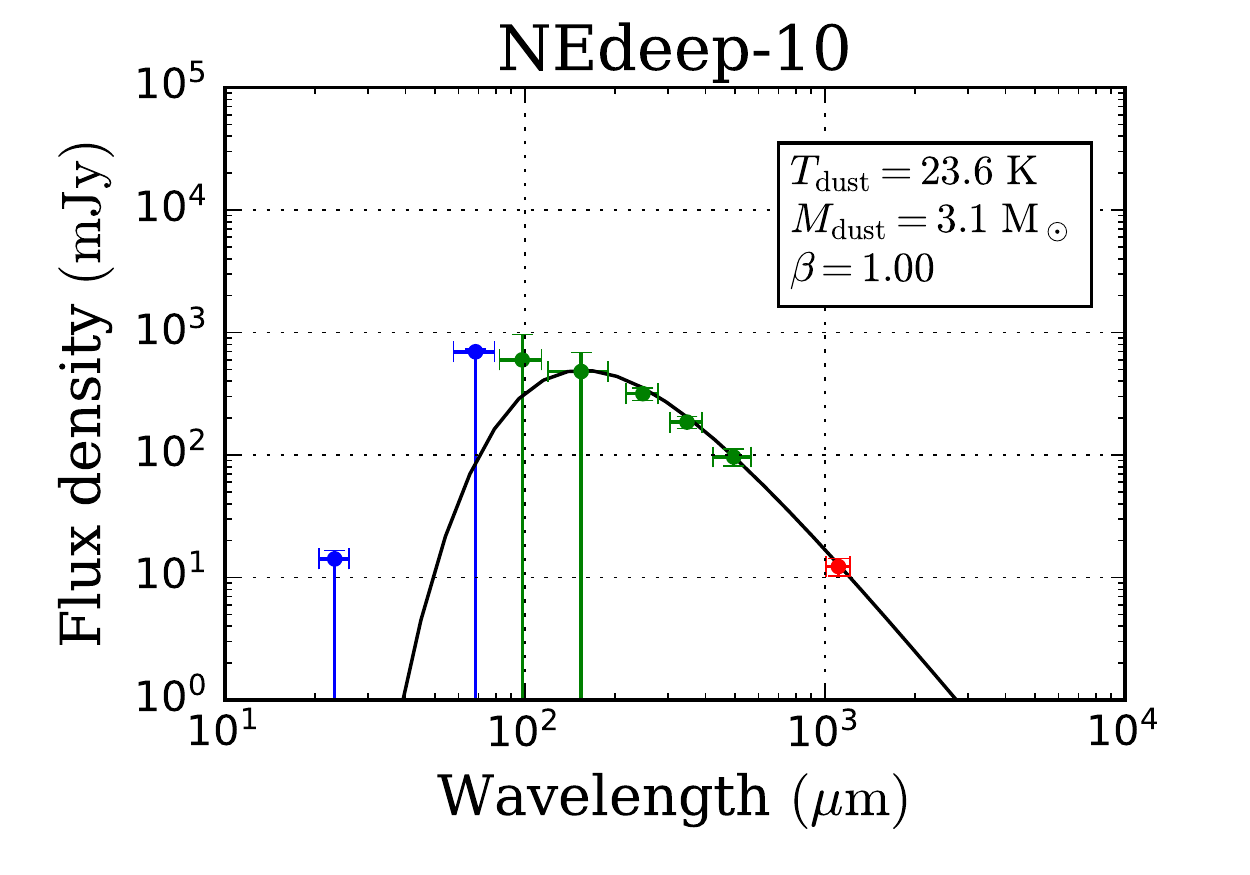}
\plottwo{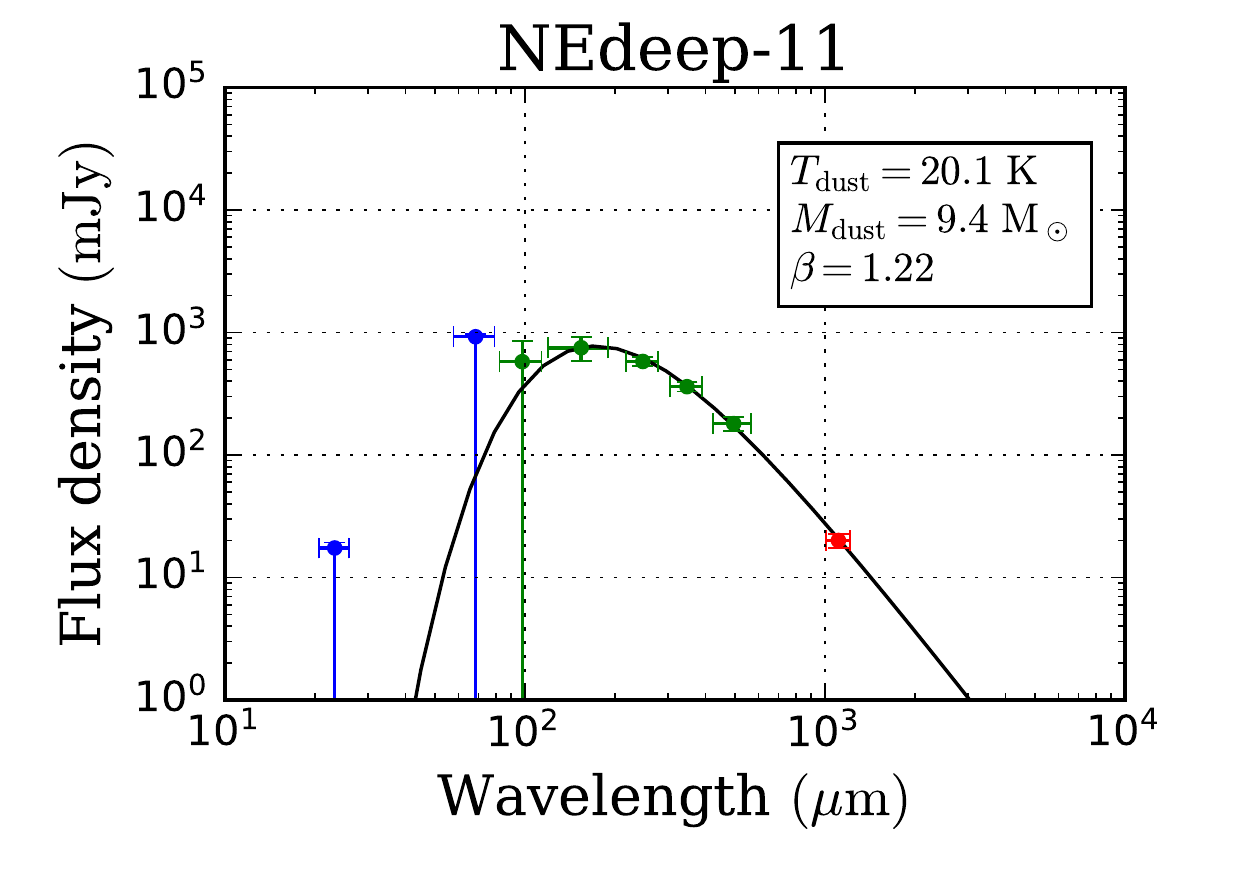}{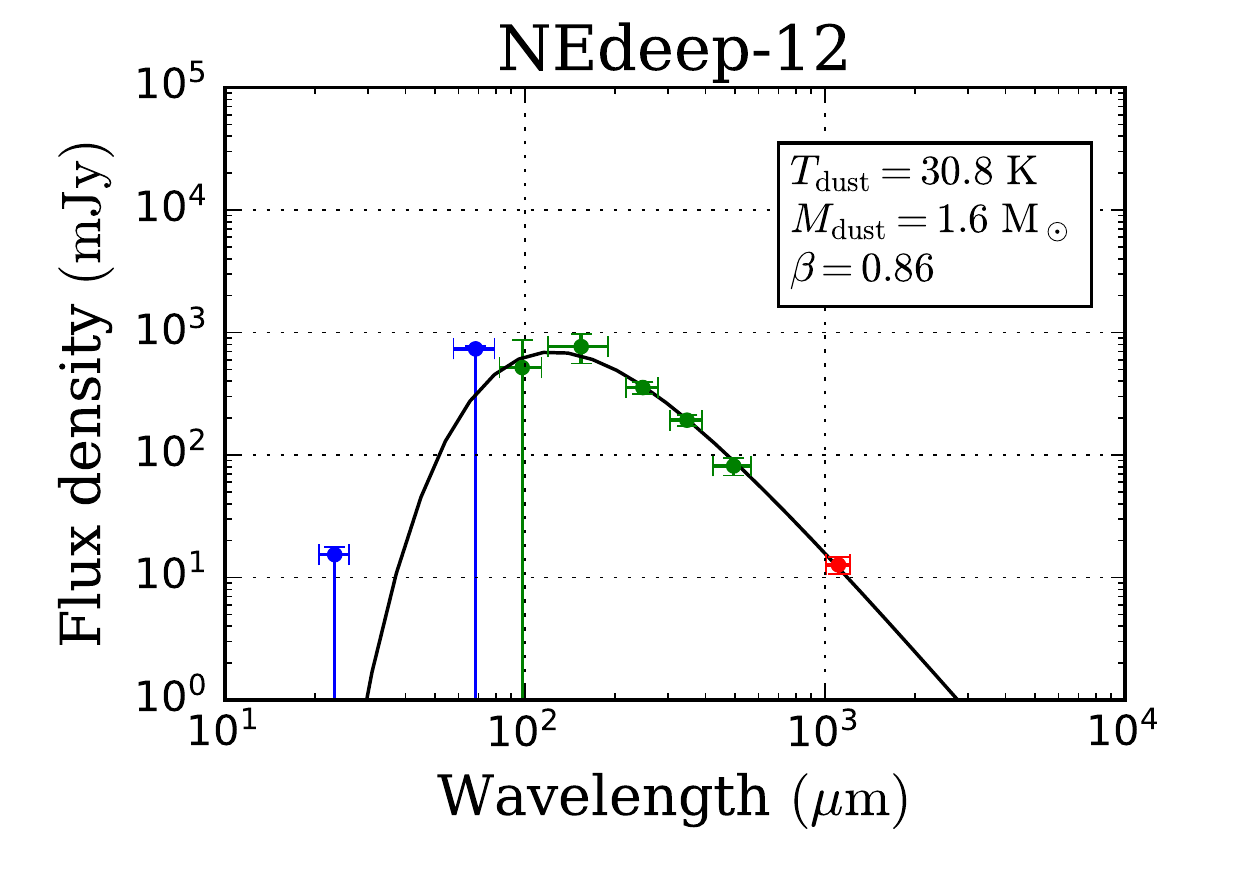}
\plottwo{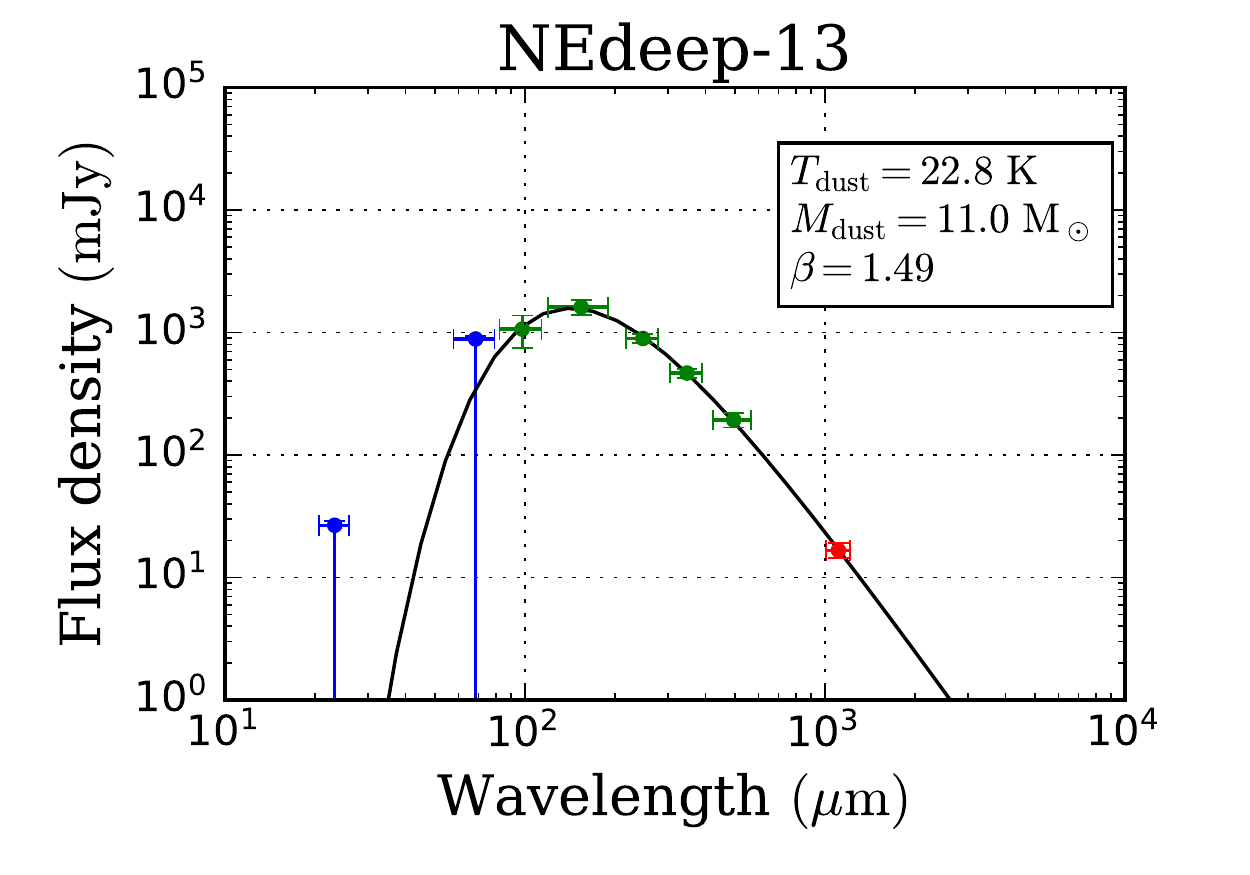}{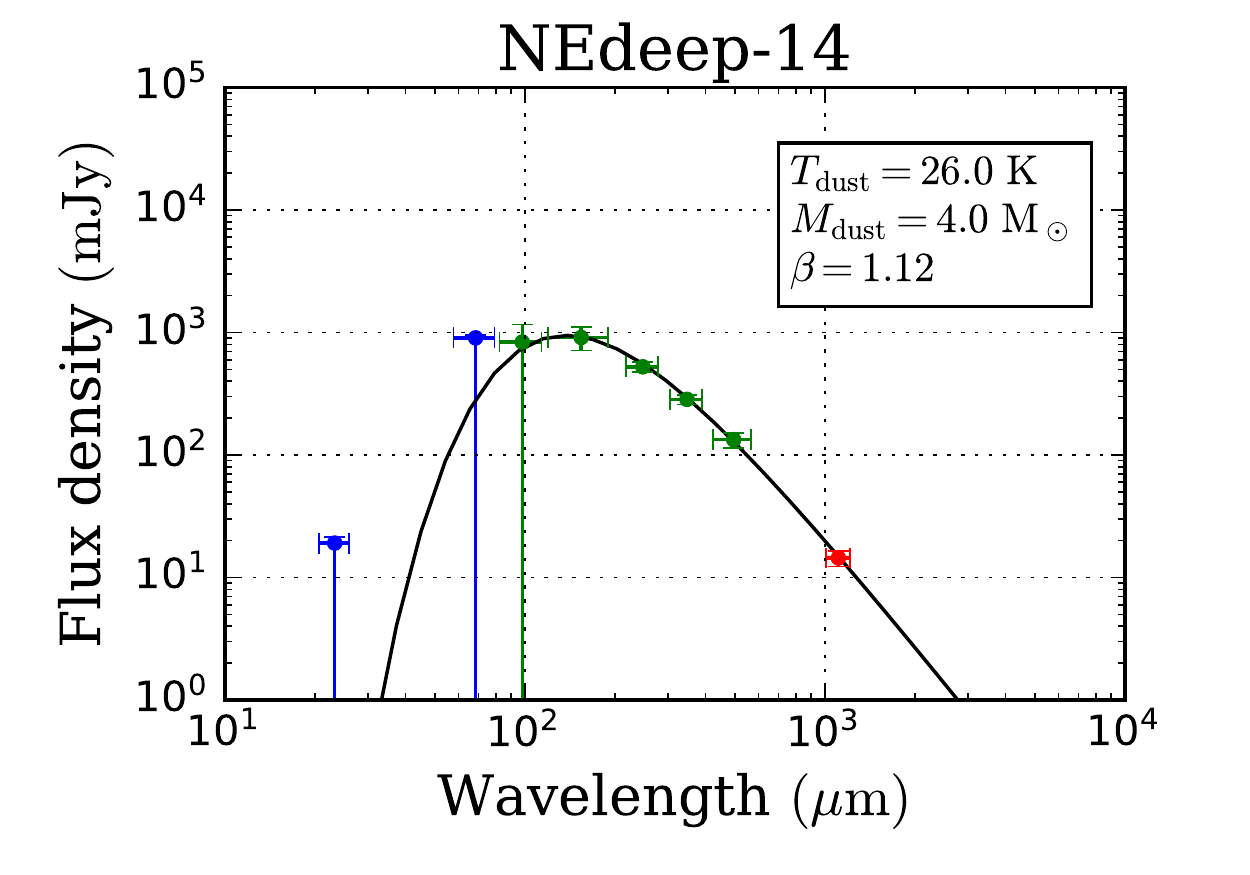}
\plottwo{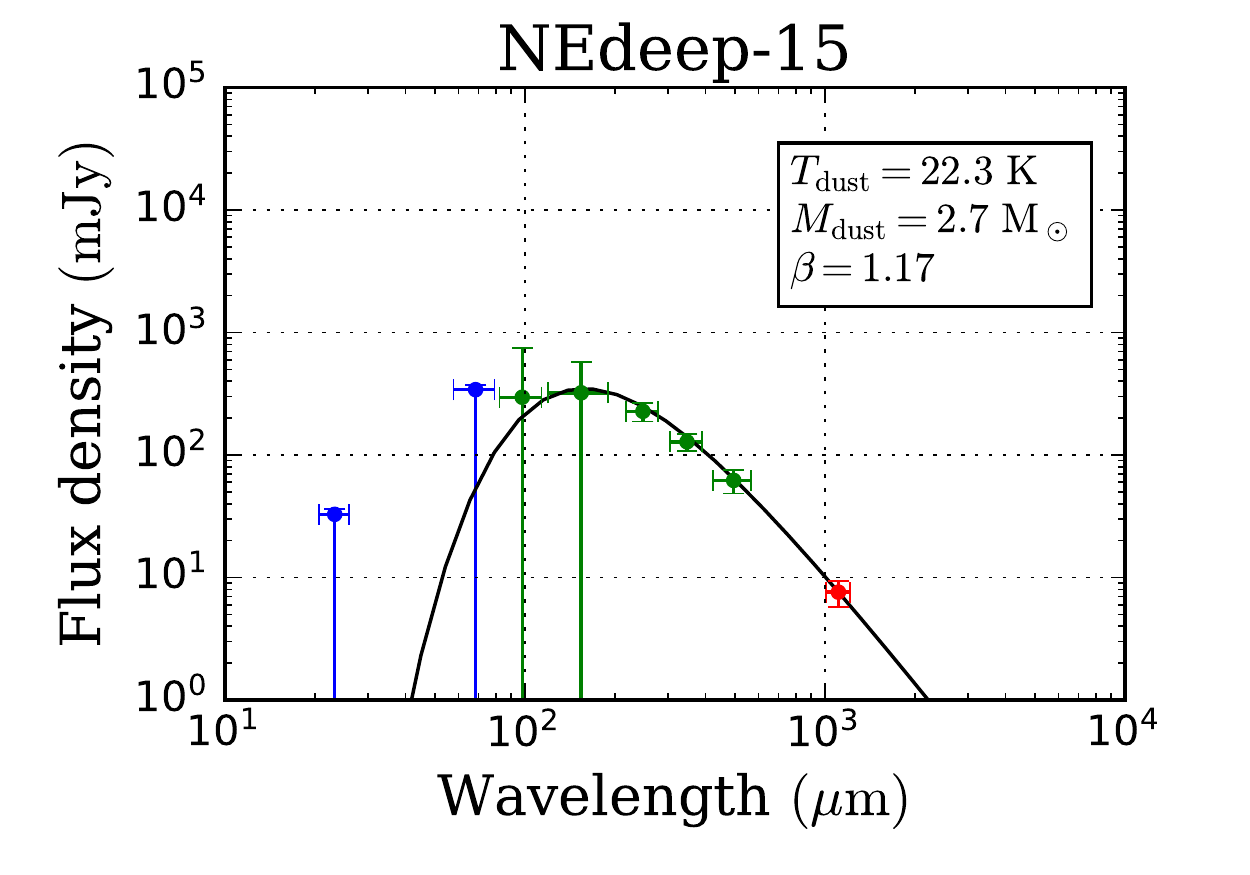}{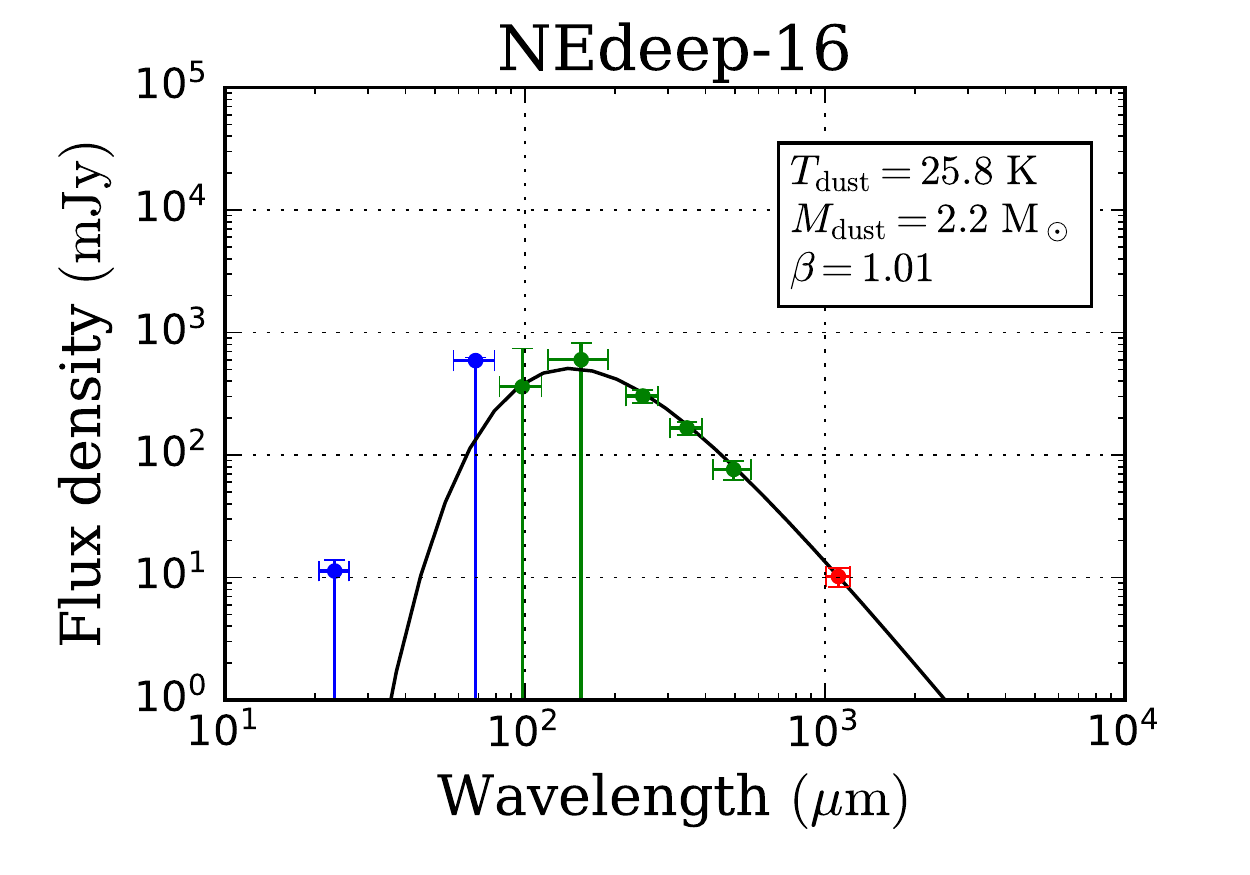}
\caption{\textit{Continued.}}
\end{figure}
\begin{figure}
\epsscale{1.1}
\plottwo{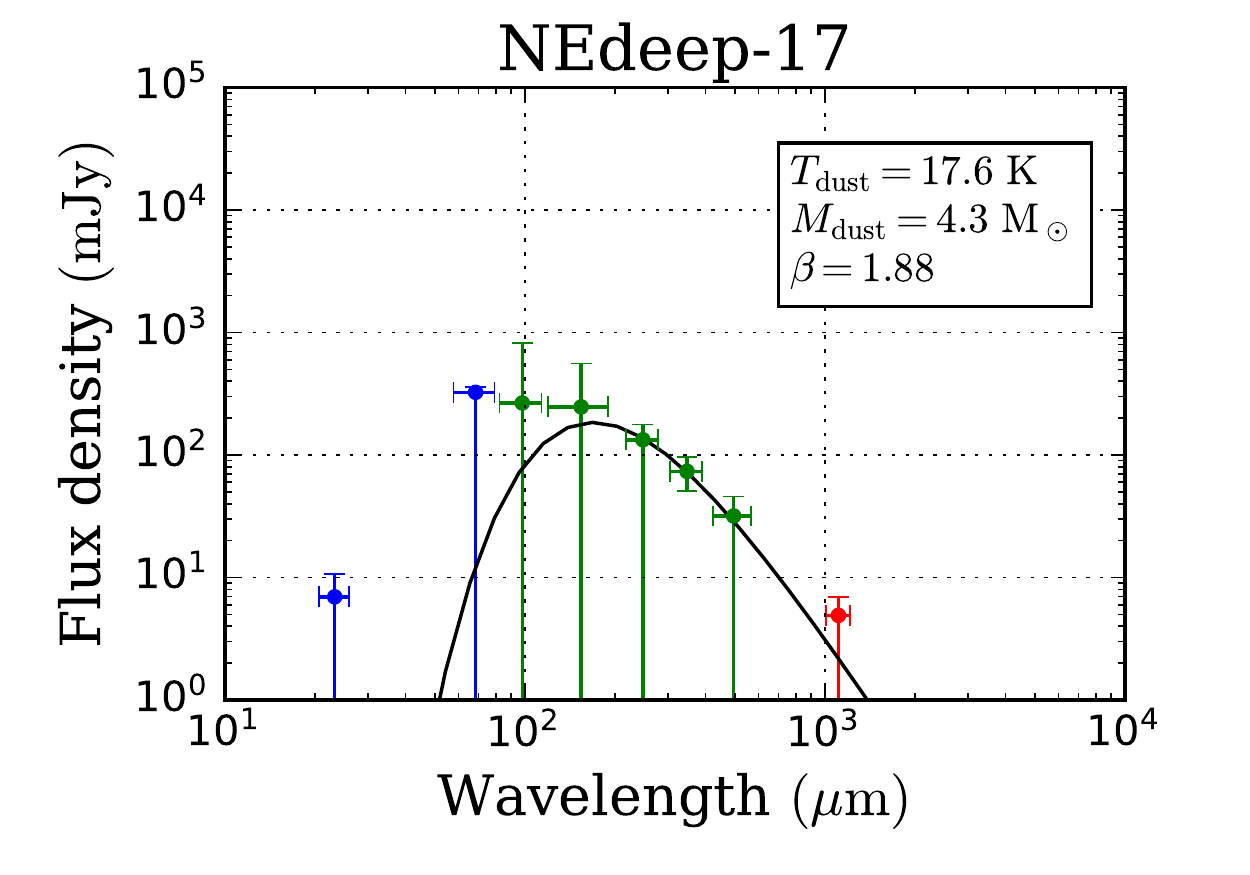}{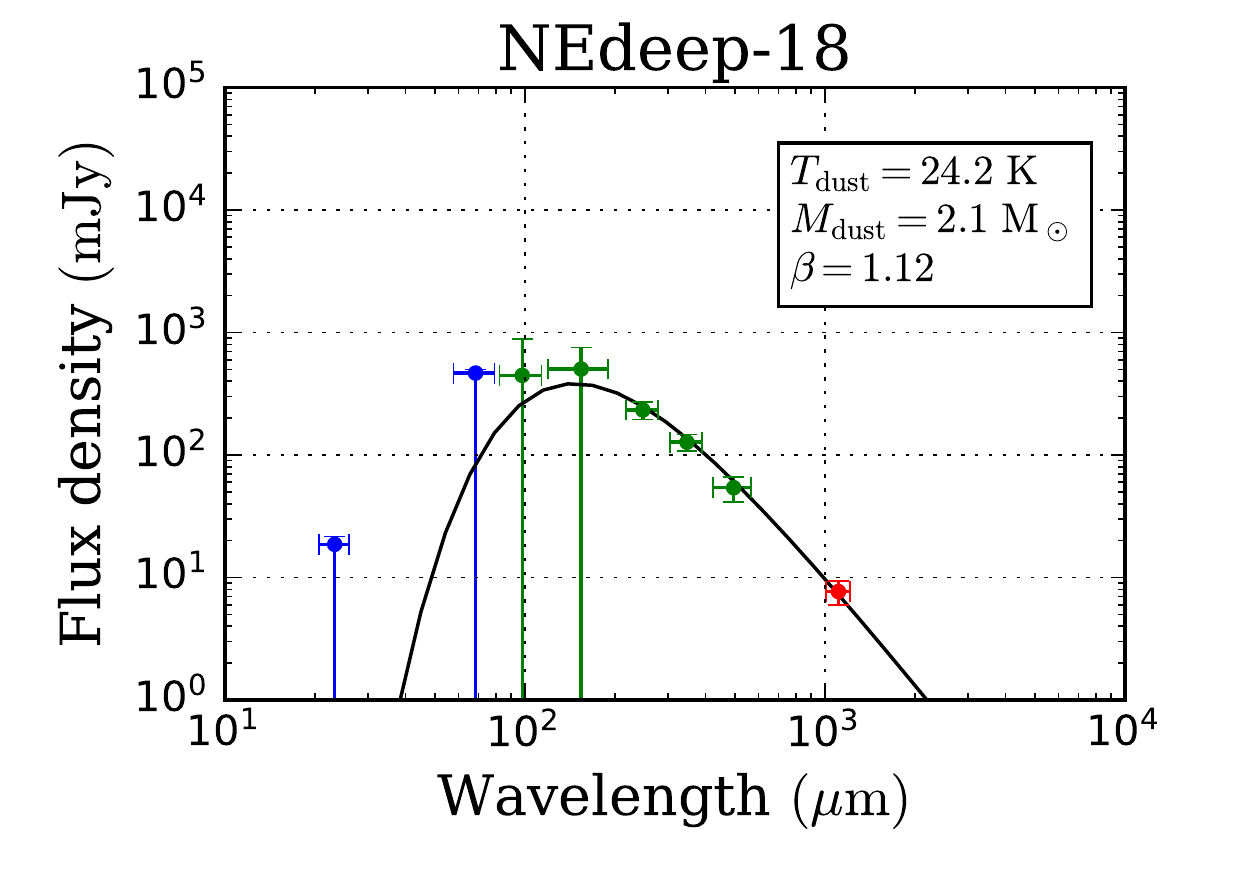}
\plottwo{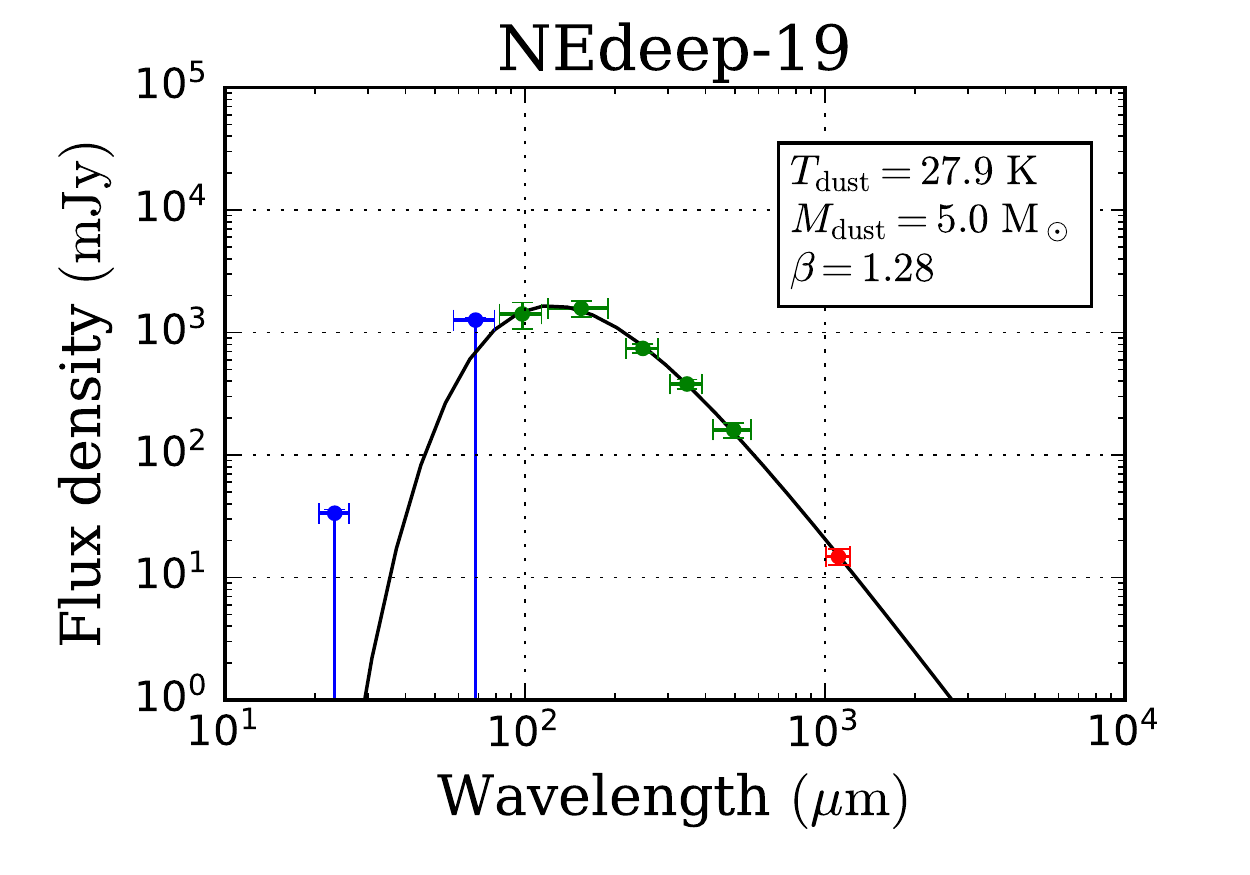}{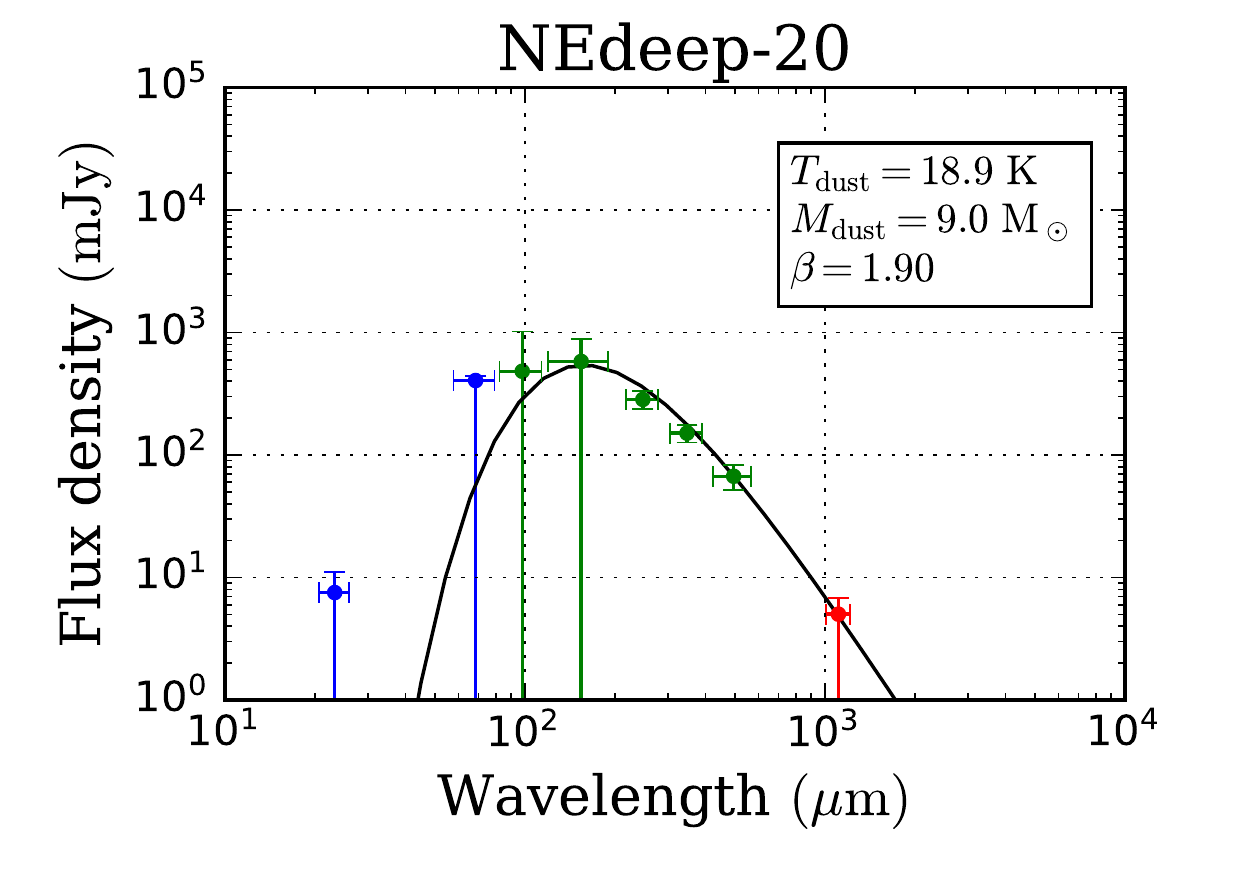}
\caption{\textit{Continued.}}
\end{figure}

\end{document}